\documentclass[sigconf,nonacm]{acmart}

\setcopyright{none}
\usepackage{booktabs}
\usepackage{listings}
\usepackage{indentfirst}
\usepackage{subcaption}
\usepackage{bm}

\usepackage{tikz}
\usepackage{xcolor}
\usepackage{listings}
\usetikzlibrary{shapes,arrows,positioning,shadows.blur}

\definecolor{terminalgreen}{RGB}{0,255,0}
\definecolor{terminalpurple}{RGB}{255,0,255}
\definecolor{terminalyellow}{RGB}{255,255,0}
\definecolor{terminalblue}{RGB}{104,113,255}

\usepackage{multirow}
\usepackage{svg}
\usepackage{graphicx}
\usepackage{algorithm}
\usepackage{comment}
\usepackage{algpseudocode}
\usepackage{booktabs}

\usepackage{amsmath}

\iftrue
\newcommand\kurt[1]{\textcolor{blue}{[Kurt: #1]}}
\newcommand\luca[1]{\textcolor{olive}{[Luca: #1]}}

\else
\newcommand{\kurt}[1]{}
\newcommand{\luca}[1]{}
\fi

\usepackage[utf8]{inputenc}
\usepackage{filecontents}

\DeclareMathOperator*{\argmin}{arg\,min} % thin space, limits underneath in displays

\tikzstyle{title} = [draw, rectangle, font=\bfseries]

\usepackage{xspace}
\title{Evaluating the Robustness of a Production Malware Detection System to Transferable Adversarial Attacks}

\newcommand\blfootnote[1]{%
  \begingroup
  \renewcommand\thefootnote{}\footnote{#1}%
  \addtocounter{footnote}{-1}%
  \endgroup
}
\newcommand{\code}[1]{\texttt{#1}}

\author{Milad Nasr$^{1\dagger}$,
Yanick Fratantonio$^2$,
Luca Invernizzi$^2$,
Ange Albertini$^2$,
Loua Farah$^2$, \\
Alex Petit-Bianco$^2$,
Andreas Terzisr$^1$,
Kurt Thomas$^2$,
Elie Bursztein$^2$,
Nicholas Carlini$^{1\ddagger}$\\ \noindent \\$^1$Google DeepMind \xspace $^2$Google}
\newcommand{\gmail}[0]{Gmail\xspace}
\newcommand{\magika}[0]{Magika\xspace}

\renewcommand{\shortauthors}{Milad Nasr et al.}
\begin{document}

\begin{abstract}

 %  \kurt{For the current title, its inaccurate to call Magika malware classification. Its a file-type detector that can be used to detect files trying to masquerade as another file type, but it has no innate detection itself. Accurate filetype detection is critical though to sending files to an optimal AV engine.} \luca{I think the current title is ok as this paper is evaluating the whole system against adversarial attacks by targeting a weak link}

    As deep learning models become widely deployed as components within larger production systems, their individual shortcomings can create system-level vulnerabilities with real-world impact.
    This paper studies how \emph{adversarial attacks} targeting an ML component can degrade or bypass an entire production-grade malware detection system, performing a case study analysis of \gmail's pipeline where file-type identification relies on a ML model.

    The malware detection pipeline in use by \gmail contains a machine learning model that \emph{routes} each potential
    malware sample to a specialized malware classifier to improve accuracy and performance. This model, called \magika, has been open sourced.
    By designing adversarial examples that fool \magika,
    we can cause the production malware service to incorrectly route malware to
    an unsuitable malware detector thereby increasing our chance of evading detection.
    Specifically, by changing just 13 bytes of a malware sample, we can successfully evade \magika in 90\% of cases and thereby allow us to send malware files over \gmail.
    %\luca{This is the result of the first attack, which doesn't check if the resulting file is broken (and thus effectively not malicious anymore). I'd report the result of the format preserving one, as it is the more significant result of the paper.}
    %
    We then turn our attention to defenses,
    and develop an approach to mitigate the severity of these types of attacks. 
    For our defended production model, a highly resourced adversary requires 50 bytes to achieve just a 20\% attack success rate.
    We implement this defense, and, thanks to a collaboration with Google engineers, it has already been deployed in production for the \gmail classifier.
    %We implemented our defense and shared it with Google, following the customs of responsible disclosure, to improve \gmail's defenses. %\luca{rephrased this not to break anonymity - also we say the same later on in the paper, so we needed to harmonize it.}
\end{abstract}

\maketitle

\section{Introduction}

%One of the main issue with adversarial machine learning literature is that they mostly focus on some hypothetical problem and there are a very few example of the these models used in actually products. As a results defenses also are considered in ideal settings. While several attempts on designing defenses for adversarial examples, however, the problem is considered unresolved. Nevertheless, we do not see any effect of that in practice. Similar attacks on language models get recently many attention but still they haven't been used to widely in practical applications. \magika has been recently introduced which replaces traditional file type detection systems. \magika uses a classification model trained on a large corpus of files. On the existing evaluation datasets \magika shows higher performance compared to the existing rule based approaches. Based on the report from Google\magikabog, 
% \magika designed to be used at scale with the goal of improving Google users’ safety by improving routing the files to the correct malicious content scanner. \magika is used in GMail, Drive, and Safe Browsing. 
\blfootnote{$\dagger$ Currently at OpenAI\\$\ddagger$ Currently at Anthropic}
Adversarial examples---evasion attacks that cause machine learning models to misclassify adversarially modified inputs---have been studied for over ten years.
Such attacks have been demonstrated on nearly every domain of machine learning, ranging from image classifiers \cite{szegedy2013intriguing,goodfellow2014explaining,madry2017towards} to speech recognition systems~\cite{carlini2018audio,alzantot2018did,qin2019imperceptible} to natural language processors~\cite{zhang2020generating,zou2023universal}.
These attacks typically work by performing gradient descent on an input example
to maximize its classification loss.

And yet, ``real attackers don't compute gradient gradients'' \cite{apruzzese2023real}: despite these attacks being known to be possible, there are very few cases where actual adversaries have performed gradient based attacks to cause some specific harm on a production system.
Why is this?
The most common answer that has been argued for several years \cite{gilmer2018motivating} is that
performing adversarial attacks is rarely the \emph{easiest} method to achieve the adversary's ultimate objective.
For example, adversaries who wish to fool image classification models could simply generate
random examples until they find a mistake \cite{gilmer2018motivating},
could perform simple modifications like increasing or decreasing the contrast or brightness,
or could make random changes to the background until one was successful \cite{hendrycks2019benchmarking}. In this work we show a real world setting where an adversary should use existing adversarial example based approach instead of other modification to misclassify the input.

%\kurt{This seems to be a valid argument still?}\milad{not for how magika was implemented it is much easier to actually run a } 
%\luca{perhaps we can simulate this: randomly flipping bits, how much time would it take to bypass magika?}
%\yanick{and to add: if it's not that easy, then we should state, otherwise one is left wondering whether that's the case for magika or not.}\milad{Figure~\ref{fig:random_flip} does this}

In particular, we study the question of how an adversary could construct malware that evades
detection by the \gmail malware detection system using adversarial approaches~\cite{pierazzi2020intriguing}.
Typically, producing an evasive malware sample requires the laborious process of trying to construct a file that leverages advanced obfuscation techniques, to be both malicious while also hiding its behavior from malware detection. % \luca{note to self: add citation to VT alternatives used in cybercrime to do exactly this}
But we find that an alternate approach is possible. 

In production environments, malware detection is slow and expensive; it is not practical to run 
the most advanced malware classifiers over every potential sample. 
And so, to improve efficiency, detection is often split into three phases. 
First, a fast signature-based classifier scans for obvious features of general malware.
Then, if nothing is detected, a second classifier \emph{routes} this sample to one (or more) specialized malware classifiers.
Third, this specialized classifier scans for more sophisticated forms of malware of a particular type (e.g., PDF malware or JavaScript malware). This optimization approach reduces the cost of system significantly lower, however, introduces a critical dependency: the security of the entire system relies on the robustness of each component, including the routing classifier.
Google, for example, uses their \magika~\cite{fratantonio2024magika} content-type classifier (in addition to more traditional content-type classifiers) ``to help improve Google users’ safety by routing Gmail, Drive, and Safe Browsing files to the proper security and content policy scanners.~\cite{magika_google_blog}''
%\yanick{Technically, magika is one of the three routing systems, the first two being file extension and traditional signature-based approach. But unclear how much of this is public.}\milad{I don't know if we should say it here or not, I leave it up to you in the experiment, I mention that we hypthosize this is the case}

We make the observation that this second classifier introduces the potential for a simple evasion attack:
by generating adversarial examples that evade the \magika classifier (in addition to evading the other existing classifiers), it is possible to completely bypass the significantly harder task of fooling the sophisticated malware classifier, by instead causing \magika to route the sample to a classifier that is not specialized to detect it.
%\yanick{should we mention that Google uses magika in addition to existing content-type detection tool? This is probably mentioned around anyways, and it would convey that these Google folks are not completely clueless about evasion possibilities.}\milad{it is mentioned a few times in the paper but we can add it here as well}
%
Because \magika is ``just'' a standard neural network designed without adversarial examples in mind  %\milad{There are two problem, first it is that the open source version is the same as the internal so even if you use existing robust approach you actually won't get much benefit also if you consider transferable attack there are not standard defenses and this is what we do as part of this paper, do you have any suggestion on what to write? },
we can apply standard techniques from the literature to evade detection.
Therefore, all that is necessary to produce evasive malware is to 
(1) evade simple signature detection, 
and then (2) fool the router to misroute the sample towards the wrong specialized classifier.
As we show, this is more easily achieved due to \magika's open-source codebase.
We note that \magika is explicitly never positioned as being robust against adversarial attacks; in fact, the authors call out this limitation and are ``looking forward to seeing adversarial examples from the community.''\footnote{\url{https://github.com/google/magika}} \cite{fratantonio2024magika}
(This paper does exactly that.)
Nonetheless, \gmail decided to add \magika as an additional and complementary file-type classifier to its critical production system, due to \magika's higher precision and recall on many critical content types. This meant that ultimately \gmail relied
on \magika's robustness for those content types. This is likely a common design pattern in many security systems, which rely on components that are not sufficiently mitigating adversarial attacks.

%\kurt{Subtle framing: Magika is never positioned as being robust, and in its paper, calls out evasion as a risk. Its purpose is value-add over trivial obfuscation (e.g., breaking a detection rule). The emphasis should be more on the defense: how can you harden a system not designed to be used in security, but that is nonetheless used by security teams, to be more robust? This is likely a common design pattern in many security systems, which rely on components that aren't in fact designed for adversaries.}\milad{one of the main issues here is this is open sourced} \luca{rephrased to capture kurt's note}

\paragraph{Contributions.}
The focus of this paper is in analyzing the \magika classifier as a case study for what can go wrong when machine learning models get integrated into larger systems: the space of the attack on a production system will include attacks on any machine learning component.
%
%\kurt{This was already the case when security teams used `file`. So the question is how to harden individual components that go into a system.}

Specifically, we show how to leverage recent advances in constructing discrete NLP adversarial examples \cite{zou2023universal}
and apply these attacks to generate evasive malware samples.
By changing just a handful of bytes in any given file, 
we show how to cause the open source \magika router to send common malware file types (e.g., \texttt{.doc}, \texttt{.pdf})
to the incorrect backend malware detectors.

We then show that these adversarial malware samples allow us to fool the production Google classifier.
To validate this attack, we work with researchers at \gmail to construct malware samples
 that we show can evade detection by the entire pipeline.

Next, we turn our attention to defenses.
We consider several techniques from the literature and develop our own refinement that
we show can significantly increase the difficulty of constructing effective transferable
adversarial examples in practical settings,
and we then work with Google engineers to deploy this classifier to production to improve users' safety.

We argue that while the vast literature has shown the difficulty in fully solving the adversarial example problem, a practical security engineering mindset is crucial for deployed systems. The goal should be to implement defenses that significantly increase the cost for the adversary, making the ML component sufficiently robust in its operational context so it is no longer the weakest link, even if theoretical vulnerabilities remain.

Finally, we conclude by discussing the relative costs and benefits of open sourcing machine learning systems used in production environments. 
While open-sourcing \magika might enable easier attacks, it also allows security researchers to study real-world scenarios and develop more effective defenses.
%This approach fosters a more complete understanding of adversarial attacks in production, moving away from idealized settings where defense mechanisms might appear impossible. 

% Rule based approaches can be fooled most of the time by hand crafted modification to the file types. There are many attacks rely such modification to confuse the defenders. One of the goals of \magika is identify the files that are misclassified by hand crafted malicious files. However, \magika is machine learning model and they are vulnerable to adversarial example. So the goal of this work is to evaluate the weakness and strength of using machine learning model in real applications and evaluate the best defense strategies.  

% For double blind submission
%\paragraph{Responsible disclosure.}
%We (the authors of this paper)
%We disclosed our attack to Google, 
%and have received advance permission to submit this paper to USENIX Security
%while the attack is being patched.

\section{Problem Statement and Background}

We begin with a review of 
malware detection,
file type detection,
adversarial examples, and
attacks on natural language processing (NLP) models.

\paragraph{Malware Detection}

%Malware is a major security concern online.
Malware includes any malicious program that subverts a user's device to engage in harmful activities (e.g., steal credit cards and other data, waste system resources to mine cryptocurrencies, send spam, and more). Security systems that detect and prevent malware from executing often rely on a defense-in-depth approach. One layer of this defense is preventing attackers from distributing malicious files via email.
%Today, millions of compromised websites use drive-by download attacks to infect vulnerable computers~\cite{skoudis2004malware}. \kurt{Might make more sense to frame this for email attachment detection, given that's what you analyze. Drive-bys are unrelated to the detection systems you explore.} \luca{We can't use a 2004 paper to say that today drive-bys are common. They aren't anymore - ransomware/spyware is more prevalent. Let's cite the mandiant report \url{https://cloud.google.com/security/resources/m-trends} or Cisco Talos: \url{https://www.cisco.com/c/dam/global/en_au/pdfs/talos-ir-quarterly-threats-2024-q2.pdf}} 
%
Cisco for example reports blocking 9 million malicious emails per hour\footnote{\url{https://www.cisco.com/c/dam/global/en_au/pdfs/talos-ir-quarterly-threats-2024-q2.pdf}},
and Google extensively applies malware detection to prevent spread via Gmail ~\cite{gmail_security_blog}.
%
%Let's cite the mandiant report \url{https://cloud.google.com/security/resources/m-trends} or Cisco Talos: } 

Given the widespread malware threat, it's no surprise that researchers have devoted considerable effort to developing methods for collecting, studying, and mitigating malicious code~\cite{madan2022tools}.
Most relevant for this paper, recent malware detection methods specialize to specific types of malware, enabling an increase in detection accuracy and reduction in processing time---something important when scanning billions of emails per day~\cite{magika_google_blog}.

\paragraph{File Type Detection.}
In order to enable filetype-specific malware detection, it is first necessary to
detect the type of the file.
Simple file type detection tools such as the \code{file}~\cite{file_command} utility work by scanning the file with basic regular expressions.
This works because most file types are designed to be easy to identify in benign scenarios, and so it is only necessary to look at the first few bytes of the header or footer. 
To evade detection by simple file type detection, prior work~\cite{albertini2015abusing} has already studied and developed effective techniques, which usually involve finding inconsistencies between the patterns in these systems and how target applications can process the files. For example, in many cases if the adversary adds a different header to the file, the applications might skip the unrecognized header and start from the known patterns while naive file type detection applications that only look at the first few bytes would not detect them. 
%In practice, many file type detection tools have capabilities to scan the files for multiple headers to prevent such attacks. 

\paragraph{\magika.}
To address the shortcomings of existing file type detection approaches, Fratantonio \emph{et al.}~\cite{fratantonio2024magika} trained a machine learning model to detect the type of any given file.
This technique, released as an open-source tool called \magika, is used at Google~\cite{magika_google_blog}. Specifically, as reported by Google, "\magika significantly boosts Google users' safety by accurately routing Gmail, Drive, and Safe Browsing files to appropriate security scanners, improving file type identification by $12\%$ (F1 score) compared to previous rule-based systems. This accuracy increase allows for 11$\%$ more files to be scanned with specialized malicious AI document scanners, reducing the rate of unidentified files to just 3$\%$."

The version of \magika discussed in \cite{fratantonio2024magika} employs a convolutional architecture for file classification. It extracts the first, middle, and last 512 bytes of a file, using their direct byte values as input to the model.\footnote{Newer versions of \magika use a slightly different architecture and parameters; such differences are not relevant in the context of this paper.} If a file is smaller than the required size, the input is padded to the appropriate size. After the bytes are processed by a standard deep neural network, the model outputs  a probability over expected content types.

While, for some content types, \magika is a strict improvement over the \texttt{file} utility in terms of benign accuracy on non-adversarial data, it was also not designed for robustness \cite{fratantonio2024magika}.
The focus of this paper is to (1) investigate the security consequences of deploying \magika in settings with adversaries, and then
(2) improve its robustness to attack so that it can be used more reliably.

\begin{figure}[t]
    \centering
    \includegraphics[scale=0.23]{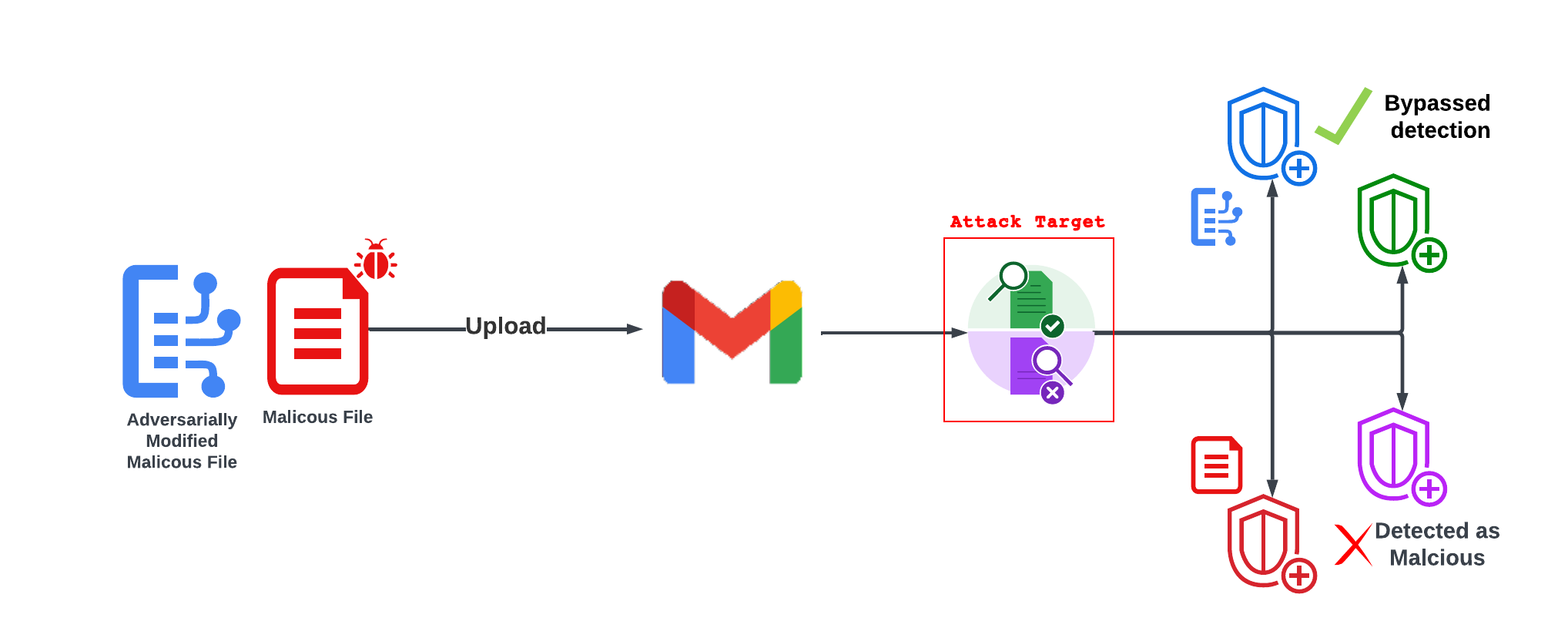}
    \caption{{\bf Overview of our attack strategy against Gmail.} An attacker aims to send a malicious file as an email attachment. The attacker adversarially modifies the file such that the \magika file-type detector run by Google's servers misroutes the file to an incorrect, specialized security scanner (e.g., a PDF is sent to a Windows executable scanner), thus evading detection. Absent these adversarial modifications, the file would be sent to the correct security scanner and be accurately detected as malicious.}
    \label{fig:gmail_overview}
\end{figure}

\paragraph{Adversarial examples.}
An adversarial example is an input designed to deceive a machine learning model into making an incorrect prediction. Adversarial examples are crafted by adding subtle perturbations to the original data so that the object remains the same (e.g., images look the same, audio sounds the same, text means the same). 
Mathematically, this can be expressed as the following:

\begin{equation}\label{eq:erm_untargted}
  \bm{x}^* = \bm{x} + \arg \min \{ \bm{z}:O(\bm{x}+\bm{z}) \neq O(\bm{x}) \} = \bm{x} + \bm{\delta_{x}}
\end{equation}
where $\bm{x}$ is the original input, $\bm{x}^*$  is the adversarial perturbation, and $O(\cdot)$ represents the model's prediction.

%One popular method to generate adversarial examples is the Fast Gradient Sign Method (FGSM). It calculates the perturbation as follows:
%\begin{equation}
%    \bm{\delta_{x}} = \epsilon \times \text{Sign} (\nabla_{\bm{x}} \loss(f(\bm{x}),y) )
%\end{equation}

%where $(\nabla_{\bm{x}} loss(f(x),y)$ is the gradient of the model's loss with respect to the input $x$, $y$ is the true label, and $\varepsilon$ controls the perturbation's magnitude.

There are many techniques to generate adversarial examples \cite{goodfellow2014explaining,madry2017towards}, depending on the norm being minimized and the domain being attacked.
Since the first adversarial attacks were developed, researchers have also attempted to develop defenses that reduce the efficacy of these attacks; unfortunately, none have managed to entirely eliminate the underlying vulnerability to adversarial attack.

%\kurt{This last paragraph needs some work or could be cut. Im not sure its relevant versus what you introduce in Section 3.}\magika is a neural network and thus susceptible to adversarial examples, many works on this area have primarily focused on image classification, with methods prioritizing minimal perceptible changes to the image by minimizing the $L_1,L_2,L_{\inf}$ norms. However, in our case, the number of bytes changed is the most important limitations. Some work has explored the $L_0$ norm, but attacking \magika is more similar to those in natural language settings, which we will focus more in the next section.

\paragraph{Generating natural language adversarial examples.}
Due to the rapid progress in the field of natural language processing~\cite{devlin2018bert,brown2020language,radford2019language},
there has been significant interest in developing adversarial examples for language models.
Early attacks were able to construct English sentences that fooled sentiment analysis systems~\cite{ebrahimi2017hotflip} by simply adding or removing whitespace, or inserting and removing a few characters~\cite{liang2017deep,jia2017adversarial}.
Subsequent attacks fooled summarization and question answering models by inserting unrelated sentences~\cite{li2020textfooler}.
But recent language models are significantly more advanced, and thus robust
to these simple attacks~\cite{shin2020autoprompt,zou2023universal}.

As a result of the fragility of early attacks, recent work has introduced stronger techniques that more reliably fool NLP systems.
We build our attack on the techniques of Zou~\emph{et~al.}~\cite{zou2023universal}, who introduce the
Greedy Coordinate Gradient (GCG) attack to generate adversarial examples over
discrete input domains with limits on the number of changes made to the input similar to our setting.
This attack was initially designed to introduce a new \emph{adversarial suffix} to the end of a given user command to a language model, but we will show that it is not too hard to adapt this attack to our setting where we in-place modify particular bytes of a file.

\paragraph{Robustness to Transferable Adversarial Examples} 
Defending against adversarial examples is a complex challenge, particularly due to their transferability between unrelated models, as demonstrated by Liu et al.~\cite{liu2016delving}. This transferability makes it difficult to definitively evaluate the robustness of defenses. While achieving complete immunity remains an open problem, several defense strategies have been explored. Tramer et al.~\cite{tramer2018ensemble} employed ensemble methods, combining predictions from multiple models to mitigate the impact of transferable adversarial examples. This work focuses on evaluating adversarial training, demonstrating its significant impact on model accuracy. Similar to what we will do, Shumailov et al.~\cite{shumailov2019sitatapatra}, develop approaches to prevent transfer attacks with a cryptography-based approach.

\section{Attack Objective and Threat Model}

%\kurt{It would be helpful to have a brief diagram of the AV detection architecture. E.g., a message comes in over email; magika is used to determine a filetype. Its then routed to the appropriate AV engine. Highlight in red which part of the system you attack. Likewise, indicate what your success metric is: is it pathing through the whole system, or just tricking the router?} 
Our attack objective is straightforward: we aim to send malware via an attachment in \gmail as shown in Figure~\ref{fig:gmail_overview}. As discussed, Gmail's malware detection consists of lightweight signature matching, followed by one or more file-type routers, one of which is called \magika, that sends samples to the most appropriate anti-virus scanner (e.g., PDF scanner, Windows executable scanner). Our attack objective focuses on the routing logic.\footnote{We assume that signature matching is bypassable by an attacker who uses a new malicious file not present in a previous attack campaign.} 
%\kurt{Re-worked the previous lines. Please verify. We probably should add a footnote for why we don't really worry about signature detection, because we assume a new sample outside a previous campaign.}
%

We assume the adversary initially constructs the malware sample without accounting for
any possible antivirus defenses, yielding some document or executable that will perform the desired behavior.
Then, the adversary takes this file and attempts to modify it so that it will successfully
evade detection by a given set of security classifiers.
Splitting the attack into two steps in this way does, in theory, make the attack ``harder'':
conceivably one could design the malware with knowledge of the detection techniques
and thus make the evasion easier.
But in practice, the expertise needed to develop effective malware is often different from that
of constructing adversarial examples, and therefore this dividing of responsibilities is,
we believe, more realistic.
We consider two possible settings of adversary knowledge to construct the evasion attack.

\paragraph{Full Whitebox}
The first threat model we consider is a \emph{full white-box} threat model, where an adversary
has complete access to the model including the exact parameters used in deployed model.
While we will show that attacks under this threat model are extremely effective,
they require the (strong) assumption that an adversary for some reason
has access to the model parameters.
Despite this, in Section~\ref{sec:attack} we show that this is a realistic
threat model for attacking \magika.

\paragraph{Black-box}
The second threat model considered in this paper is that of the
black-box adversary who still
has the same attack objective, but where the defender is allowed to keep a relatively small
set of secrets that defend the system.
This is a common practice in many security applications (e.g, secret-key cryptography,
address space layout randomization~\cite{pax2001aslr}, stack canaries, or pointer authentication).

In Section~\ref{sec:defense}, we evaluate several techniques to mitigate
\magika's vulnerability to attack under this threat model.
Limiting the adversaries knowledge in this way does not completely prevent all forms of attack.
Specifically, two general attack strategies remain:
(1) \emph{query attacks}~\cite{chen2017zoo,brendel2018boundary} make oracle queries to the deployed system in order to understand the way the deployed classifier
behaves on specific inputs, and adapts the attack accordingly;
and (2) \emph{transfer attacks}~\cite{papernot2016transferability} generate white-box adversarial examples on one classifier, 
and replay them on another classifier in the hope they will remain effective.

\subsection{Distortion metrics}
Throughout this paper we evaluate our attacks based primarily on whether or not they succeed.
But because we can eventually achieve 100\% success rate in every instance for undefended models %\luca{isn't this by definition? if you change every bit, the original file can become anything you want. Maybe we want to say that we achieve that 100\% changing fewer than N bytes?}\milad{we discuss in distorsion },
we additionally break down the success rate of our attack as a function
of the number of bytes changed in the malware sample.

In some machine learning settings, measuring distortion as ``number of changes introduced''
(e.g., number of pixels perturbed for an image classifier, or number of words changed
for a natural language processing model) makes sense.
There are two reasons for this:
\begin{itemize}
    \item First, an image or natural language sentence
    is only an adversarial example if it would still be labeled by
    a human as the original label; thus, the human acts as the oracle who decides the
    label of the image or sentence.
    \item Second, most images or text are shown to humans and machines simultaneously,
    and so introducing extreme noise would, if not be detected as malicious by the machine,
    be detected as malicious by the human.
\end{itemize}

But in the space of malware, neither of these motivations are relevant as demonstrated by polymorphic strains of malware where attackers produce arbitrarily diverse files via unpacking, encryption, or functionally-equivalent code snippets. % \kurt{You could cite the long tradition of polymorphism in malware to evade detection. Your approach is just another evolution of this.}
Humans are not the judge of whether or not a file is malicious or not;
it either performs malicious behavior or it doesn't.
If, having modified 1000 bytes, the sample is still malicious, then it is still a valid malware sample.
And further, there is no inherent reason why attacks that change 100 bytes of a file are
any ``more obvious'' than attacks that only changed 10 bytes---modifying the bytes within an HTML comment, no matter how many, does not alter the visual appearance of the rendered web page.
%
%There is no on-line human who will be inspecting the bytes of every
%sample about to be executed on a computer, and so avoiding detection
%by the human is unimportant \luca{I'd consider removing this sentence because it's somewhat redundant with the one above, and some 1-byte changes are very visible to a human - I can imagine reviewers calling that out}.
%

For both of these reasons, measuring attack success rate as a function of the
number of bytes perturbed is not necessarily an optimal metric. But we nevertheless believe it is still useful.
For one, measuring the number of bytes perturbed allows us to quantify the
``difficulty'' of the attack---as has been shown in past research on adversarial
malware, it is easier to evade a malware classifier by changing a larger fraction of the
bytes in a file than a smaller fraction.
And so while it is not important that we make this number small,
the fact that it can be small indicates the inherent vulnerability of these models.
Additionally, measuring the number of bytes helps to
compare defense approaches.%\luca{broken sentence}

\section{Attacking \magika}
\label{sec:attack}
%\kurt{Small nit, but this section is very long, despite being what I assume you consider to be straight forward? I'd expedite this discussion to get to defenses.}

\begin{algorithm}[t]
\caption{Greedy Coordinate Gradient}
\label{alg:gcg}
\begin{algorithmic}
\Require Initial prompt $x_{1:n}$, modifiable subset $\mathcal{I}$, iterations $T$, loss $\mathcal{L}$, $k$, batch size $B$
\Loop{ $T$ times}
    \For{$i \in \mathcal{I}$}
        \State $\mathcal{X}_i := \mbox{Top-}k(-\nabla_{e_{x_i}} \mathcal{L}(x_{1:n}))$ \Comment{Compute top-$k$ promising token substitutions}
    \EndFor
    \For{$b = 1,\ldots,B$}
        \State $\tilde{x}_{1:n}^{(b)} := x_{1:n}$
        \Comment{Initialize element of batch}
        \State $\tilde{x}^{(b)}_{i} := \mbox{Uniform}(\mathcal{X}_i)$, where $i = \mbox{Uniform}(\mathcal{I})$  \Comment{Select random replacement token}
    \EndFor
    \State $x_{1:n} := \tilde{x}^{(b^\star)}_{1:n}$, where $b^\star = \argmin_b \mathcal{L}(\tilde{x}^{(b)}_{1:n})$ \Comment{Compute best replacement}
\EndLoop
\end{algorithmic}
\end{algorithm}

We now turn our attention to the technical question of how we can
attack \magika to make it
mistake files of one type for another type.
Once this is achieved, \magika will then route the malicious sample to the wrong
specialized malware detector, thus degrading detection. We focus on evading \magika  as the measure of success in most of the work.  We will also demonstrate a proof of concept that showcases the evasion of the entire malware detection system. 

%\kurt{This might be a strong assumption; ``degrading detection''? The default fallback scanner may still succeed.}

\begin{figure}
    \centering
    \includegraphics[trim=2.8cm 16cm 1cm 7cm,clip,scale=.65]{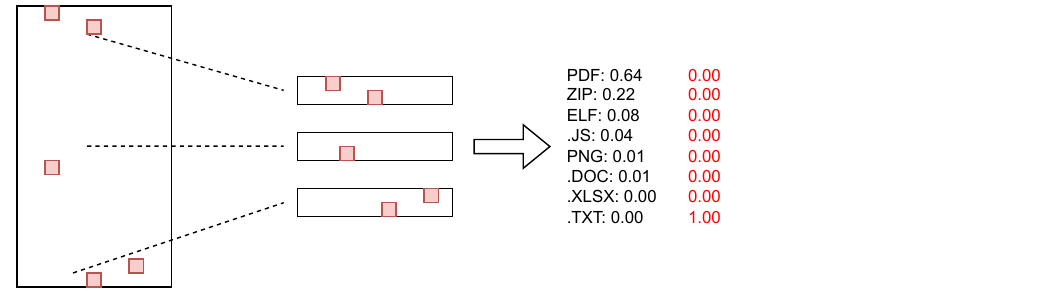}
    \caption{\textbf{Schematic of our attack.}
    \magika processes an arbitrarily large file by first
    extracting just 1536 bytes (the first, middle, and last 512 bytes),
    and then classifying these 1536 bytes with a neural network.
    By modifying just five of these bytes, 
    we will show how to successfully cause the classifier to assign
    arbitrary incorrect file type to files,
    and thus route the malware sample to the wrong specialized classifier.}
    \label{fig:attack_scheme}
\end{figure}

We hypothesize that Google used the same version as the open sourced implementation in their product. We were able to confirm this by sending several samples through Gmail and observing the behavior is similar to the open sourced version. (As we will discuss in the ethics section, we obtain advance permission from Gmail and ensure no malware is transmitted to any other user than ourselves.)
Because the open source implementation of \magika shares the exact
same set of parameters as the production version of this classifier used in
\gmail (at the time of disclosing the attack), 
we can implement an entirely white-box attack. Please note that no real users were ever targeted in this work, we will provide additional details in the ethical section of the paper.
\footnote{As a result of our experiments, and the defenses we will evaluate in
Section~\ref{sec:defense}, the version of \magika used in production is no longer the same as the open-source implementation.}
To a large degree, our attack here is a straightforward application of GCG~\cite{zou2023universal},
but with one key change:
instead of searching for a suffix that can be appended to a paragraph
to change the classification, we search for which bytes can be in-place modified to change the classification.
The bytes we modify are so-called \emph{blind spots}~\cite{brodin2023}: 
bytes of a file that can be modified without disturbing the structure of the file and still can be parsed as the correct format and maintain the malicious functionality.
The key metric in this attack that measures the attack difficulty is the number of bytes modified, 
and not the specific values they are changed to.
Hence, our focus is primarily on $l_0$ attacks that aim to minimize the number of bytes 
modified. Algorithm~\ref{alg:gcg}  summarizes our attack for attacking \magika in white-box setting. 
%In the blackbox setting instead of the computing gradient we replace the gradient with random values. 
The schema of the attack is also provided in Figure~\ref{fig:attack_scheme}.

\subsection{Experimental Setup}\label{sec:exp_setup}
\paragraph{Dataset}

We evaluate our attacks and defenses on 1,130 files, sampled by selecting 10 random test‑set data points from each of the Magika’s 113 file types supported at the time of writing. The dataset was provided directly to us by the authors of Magika~\cite{fratantonio2024magika}. This same set of 1,130 files is used across all experiments throughout the paper.

In addition to the evaluation dataset, we followed the data collection methodology described in the Magika paper to gather a dataset (20M files) from a similar distribution (VirusTotal~\cite{vt} and Github). This  dataset was used to train our substitute models for evaluation and transferability experiments.

\paragraph{Implementation}

We reimplemented Magika in JAX, initializing the model with the provided Keras weights, and verified that our implementation performs equivalently to the original. We also implemented the alternative architectures and various defenses discussed in the following sections in JAX, using data distributions similar to those used in the original evaluation.

For transfer attacks, we followed the standard approach in adversarial machine learning: the attacker trains a substitute model  on a dataset drawn from a similar distribution. The substitute models differ from the target Magika model in initialization and data shuffles.

\paragraph{Preprocessing}
Unless stated we follow Magika’s default pipeline, each file is represented by a 1,536‑byte sequence formed by concatenating three fixed‑length slices: the first 512 bytes, a 512‑byte segment sampled from the middle, and the last 512 bytes. 

\paragraph{Ensemble variant of GCG.}
In experiments where the attacker has access to multiple substitute models,
we adapt the Greedy Coordinate Gradient (GCG) attack (Algorithm~\ref{alg:gcg_blindspot}) to optimize perturbations jointly across all models in the ensemble.
Given a set of $N$ substitute classifiers $\mathcal{E} = \{f_{\theta_1}, \dots, f_{\theta_N}\}$,
we compute the gradient of the average loss over all models at each iteration.
Candidate byte modifications are scored according to this ensemble loss.

\subsection{Warm up}
To begin, we evaluate how vulnerable \magika is to adversarial examples without
considering real-world constraints.
For example, many file types have magic bytes that must never be changed; 
or have checksums that must be valid for the file to correctly process.
For the moment, we disregard these real-world constraints in order to
develop an understanding of the vulnerability of the classifier in isolation.
Then, in the following the section, we will re-introduce these ``problem space'' constraints \cite{pierazzi2020intriguing}
and show the attack remains effective.

\paragraph{Attack Procedure.}
\begin{algorithm}[t]
\caption{Greedy Coordinate Gradient  with Blind Spot}
\label{alg:gcg_blindspot}
\begin{algorithmic}
\Require Initial bytes $x_{1:n}$, blind spot positions $\mathcal{B}$, iterations $T$, loss $\mathcal{L}$, $k$, neighbor search size $B$
\Loop{ $T$ times}
    
            \State $\mathcal{X}_i := \mbox{Top-}k(|\nabla_{e_{x_i}}| \mathcal{L}(x_{1:n}))$ \Comment{Compute top-$k$ promising substitutions}
    \For{$b = 1,\ldots,B$}
        \State $\tilde{x}_{1:n}^{(b)} := x_{1:n}$
        \Comment{Initialize element of search batch}
        \State $\tilde{x}^{(b)}_{i} :=  \mbox{Uniform}((\nabla_{e_{x_i}}[\mathcal{X}_i]< 0)? [0,x_{1:n}[\mathcal{X}_i]): (x_{1:n}[\mathcal{X}_i],255])$  \Comment{Select random replacement token}

    \State $x_{1:n} := \tilde{x}^{(b^\star)}_{1:n}$, where $b^\star = \argmin_b \mathcal{L}(\tilde{x}^{(b)}_{1:n})$ \Comment{Compute best replacement}
    \EndFor
\EndLoop
\Ensure Optimized prompt $x_{1:n}$
\end{algorithmic}
\end{algorithm}

The adversary receives a malware sample as input, and has access to the weights of the \magika system in order to compute the gradients of the classification loss with respect to the input bytes. With this access, the attacker runs Algorithm~\ref{alg:gcg_blindspot}: first, identifying which input bytes have the greatest impact on the model's classification, then, selecting the $k$ positions with the largest gradient values, and finally, modifying each of these positions individually modified with all possible values. The adversary takes each of these candidates, evaluates the classification loss for each modified position and value, and ultimately chooses the modification that results in the lowest loss.

\paragraph{Evaluation.}

\begin{figure}[t]
    \centering
    \includegraphics[scale=0.6]{{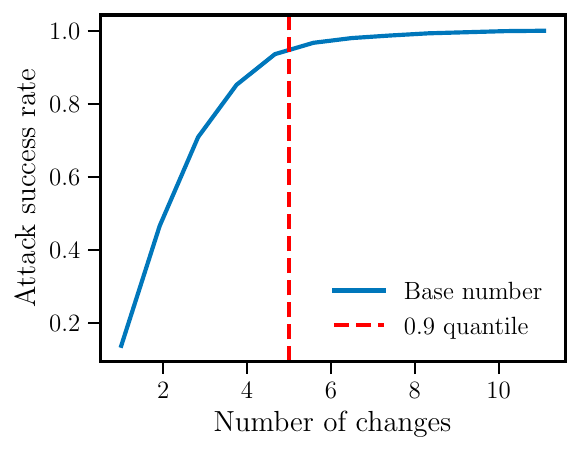}}
    \caption{Attack success rate as a function of the number of bytes modified.
    By modifying just 5 bytes, our attack can cause $90\%$ of malicious files to be
    misclassified by \magika.}
    \label{fig:direct_magika_results}
\end{figure}

 We report the average attack success rate over all of the file type in Figure~\ref{fig:direct_magika_results}. We find that \magika will misclassify nearly every category of file with just a few edited bytes.

%\luca{since it's a small set of files, I'd list the specific file types you used}\milad{no this is all 113 file types and 10 random files for each}

%While measuring the number of bytes necessary to cause a file to be misclassified is
%not a directly meaningful (recall, as we argue earlier, there are
%probably not any adversaries who can modify 5 bytes arbitrarily but can't modify 7),
%this measurement helps to establish some overall difficulty of attacking the classifier.
%
%If it was necessary to modify most of the bytes in the file, 
%then we might conclude that attacks would be challenging.
As we can see in Figure~\ref{fig:direct_magika_results},
it is straightforward for an adversary to cause \magika to misroute any given file type.
%
%This is not surprising given the extensive literature on the
%existence of adversarial example on machine learning models,
%we were surprised by how how brittle this classifier was. \kurt{Why? Nothing in its paper addresses robustness.}
%\luca{Since this initial attack does not preserve functionality of the files, it's not surprising that the classifier classifies the files as a different type. E.g., changing the first couple of bytes is a surefire way to break many files. I'd move comments about brittleness to the section where you attack using blind spots, as this section is just prelimiary work that does not validate that the attack produces valid files of the same file type}
%
Over half of binaries can be misclassified with fewer than three bytes modified;
and nearly all can with just ten. 

A few-byte modification is a nearly infinitesimal modification:
many of our files are several megabytes
and so a three byte modification represents just $0.00001\%$ of the file.
However, the reason this perturbation is so small is that \magika only inspects
a relatively small fraction of the bytes in the file;
specifically, recall that \magika samples the first, middle, and last 512 bytes of the file.
Therefore, while it is true that the \emph{absolute} magnitude of the perturbation
is exceptionally small, the magnitude of the perturbation relative to the input size
to the classifier is in line with other adversarial attacks at a ratio between 0.1-1\%.

We believe the difference between these two quantities highlights the difference between attacks as measured in the 
real world and attacks looking only at the machine learning model in isolation.
If the classifier is used in such a way that it only inspects a negligible fraction
of the file, then an attacker can completely disregard any bytes not processed
by the classifier.
In later sections, we will discuss defenses including approaches that randomize
the bytes which are scanned to improve the robustness of this classifier to attack. 
%\kurt{Something to keep in mind: the low number of bytes considered is intended to minimize scan time over a file. The selection of 3x512 was optimized for the minimum bytes to enable accurate detection in non-adversarial settings.}\milad{yes it actually might be interesting specifally for the files that gmail cares about}

\subsection{Format Preserving Attack}

\begin{figure*}
    \centering
    \begin{subfigure}[b]{0.3\textwidth}
        \centering
        \includegraphics[scale=0.45]{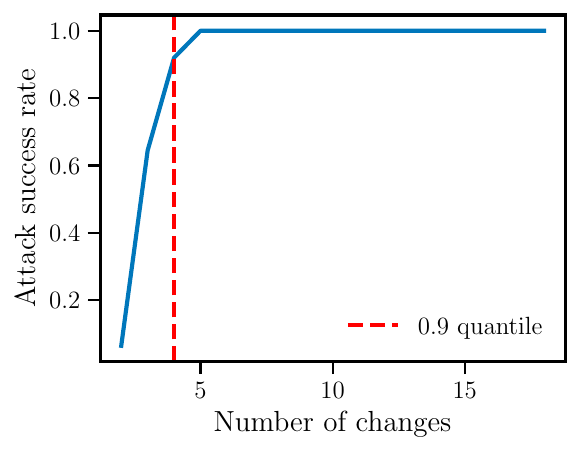}
        \caption{PDF}
    \end{subfigure}
    \hfill
    \begin{subfigure}[b]{0.3\textwidth}
        \centering
        \includegraphics[scale=0.45]{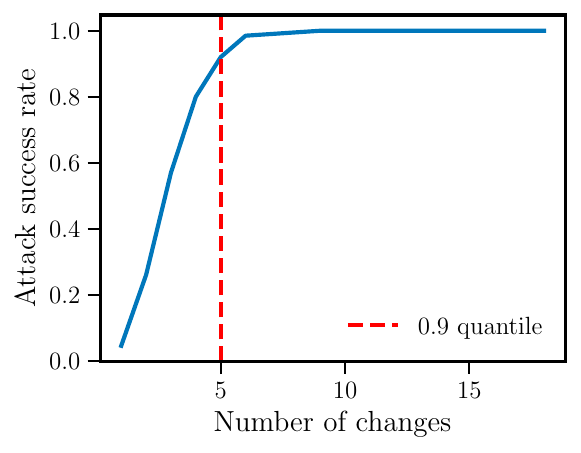}
        \caption{Zip}
    \end{subfigure}
    \hfill
    \begin{subfigure}[b]{0.3\textwidth}
        \centering
        \includegraphics[scale=0.45]{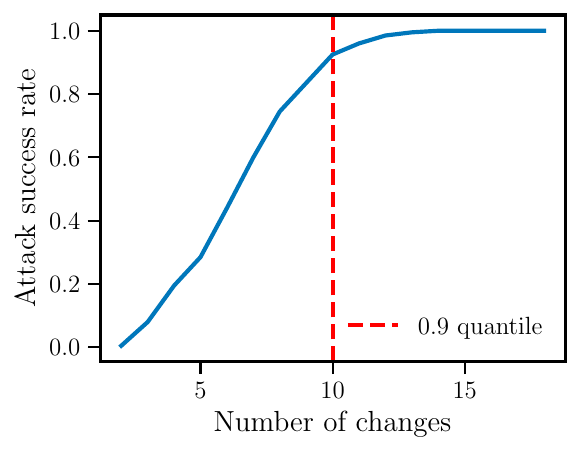}
        \caption{Docx}
    \end{subfigure}
    
    \begin{subfigure}[b]{0.24\textwidth}
        \centering
        \includegraphics[scale=0.45]{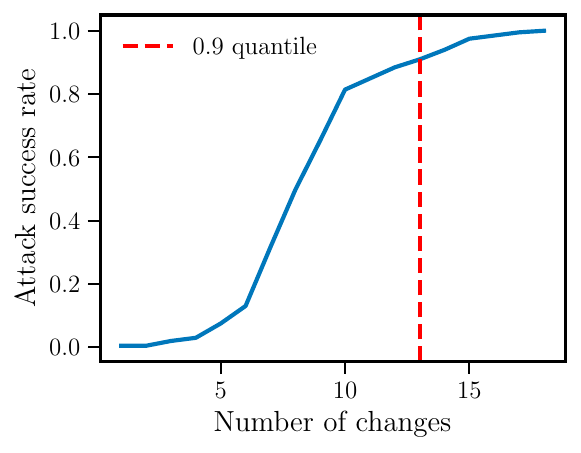}
        \caption{Xlsx}
    \end{subfigure}
    \hfill
    \begin{subfigure}[b]{0.24\textwidth}
        \centering
        \includegraphics[scale=0.45]{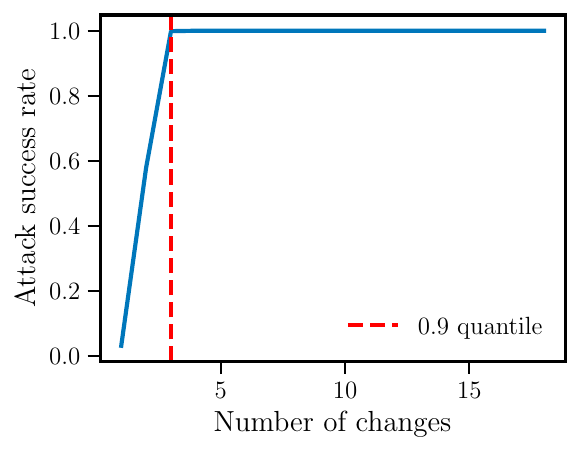}
        \caption{ELF}
    \end{subfigure}
    \hfill
    \begin{subfigure}[b]{0.24\textwidth}
        \centering
        \includegraphics[scale=0.45]{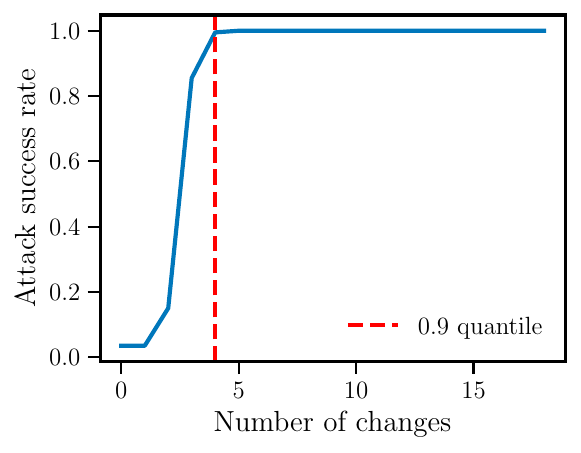}
        \caption{PNG}
    \end{subfigure}
    \hfill
    \begin{subfigure}[b]{0.24\textwidth}
        \centering
        \includegraphics[scale=0.45]{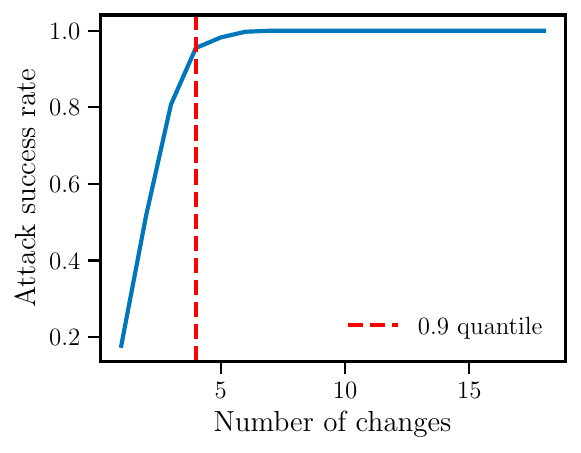}
        \caption{JavaScript}
    \end{subfigure}
    
    \caption{The cumulative number of changes required to the classify each file format file preserving the format.}
    \label{fig:magika_format}
\end{figure*}

We now adjust the prior attack to construct malware that works in the problem space,
``wherein the challenge lies in modifying real input-space objects'' \cite{pierazzi2020intriguing}.

There are two types of constraints we must satisfy in order for a file to remain valid:
\begin{itemize}
    \item \emph{Operating system constraints.} The operating system must believe that the
    file is of the type intended by an adversary, otherwise when the victim receives the
    malware and attempts to open it, the operating system will refuse to do so.
    Each operating system has a different (often exceptionally simple) file-type classifier
    that, e.g., looks at the first bytes of a file to check the presence of so-called ``magic bytes''.
    %signature = %PDF-x.y where x.y in [1.0, 1.1, 1.2, 1.3, 1.4, 1.5, 1.6, 1.7, 2.0]
    For example, PDF files typically must begin with bytes like \%PDF-1.5 in order to be valid. A naive file type detector looks at the beginning of the file and if they see \%PDF-1.5 they identify them as PDF which then can be used by the operating system to use a PDF reader to open the file. 
    But it is not sufficient for the OS to choose to open the correct program.

    \item \emph{File-specific constraints.} Nearly all file types have a much more restricted
    structure that must be valid for the file to actually run.
    For example, many file types contain metadata information (like the size of buffers or pointers to other data).
    Other file types expect a given structure (e.g., XML-based file formats would not parse if an adversary broke the XML tags).
\end{itemize}

An attack that causes \magika to mis-identify one file type as another
but breaks either the OS or filetype constraints is invalid.
In this section we develop attacks that are effective taking these constraints into consideration.
Thus, in contrast to many scenarios where adversarial examples primarily focus on designing attacks that deceive the classifier alone, this section emphasizes on preserving the file format. 
For instance, in the PDF example above, we now require that any modifications must ensure that the resulting file remains a valid PDF.
This means that crafting effective attacks necessitates a thorough understanding of the file format in question. 

\subsubsection{File types}
We evaluate the practicality of our attack on several popular file types that are commonly used to transfer malicious files, with the understanding that any other file type will likely behave qualitatively similar to these case studies.

To modify file contents while preserving the functionality, we need to find blind spots~\cite{brodin2023}, parts of the file that can be freely modified without altering the functionality of the file. File formats have different characteristics that make them more or less difficult to modify~\cite{keycom22}. 
For this reason, we make use of the  Mitra\footnote{\url{https://github.com/corkami/mitra}} tool to quickly identify
bytes that are able to be modified for 40 different file formats.
%
% Without any knowledge of the file format, Mitra can be used to inject an arbitrary payload with the {\tt -f} parameter to force a format-less content.
%
% To inject 2Kb into your input file~:\\
% \verb|mitra.py <inputfile> /dev/null --pad 2 -f|
%
From a blind spots' perspective, file formats fall in three categories:

\paragraph{Full Control} 
Some file formats tolerate to be present at any offset in the first megabytes of the file, allowing for unlimited control right from the beginning: typically, archives such as ZIP, Rar, 7zip, and hardware images such as ISO and Microsoft VHD (if the boot sectors aren't used, as the file systems are typically defined in further sectors). For these formats, an adversary can typically modify all of the first 2048 bytes of the file.

\paragraph{Parasite}
Most file formats (e.g., executables, documents, media format)
can tolerate comments or extra metadata.
%The approach differs with file formats and may require very unique manipulations to (ab)use specific characteristics of the format.
%If the format doesn't have a specific structure to be abused, maybe the file can be converted to an alternate format with identical functionality that has one.

%For example, a Mach-O executable can be included in a multi-architecture binary with a single architecture.
%For a Portable Executable, wipe most of the DOS header, the DOS stub, the Rich Header, and relocate the PE header further in the file. ELF files that are Position Independent Executable can also have their Program Headers relocated further. 

For example, GIF images have no room in their headers, so instead an adversary can make the header smaller by relocating the global palette to each frame, and abuse the first frame comments instead. Some of these configurations may be less compatible under specific environment, and some checksums may need to be updated (for TAR, PNG or archives), but this will give typically most (at least 2000) controllable bytes of the first 2048 bytes of the file for the following formats. \footnote{Other examples of parasite files are executables (PE, Mach-O, ELF, APK, JAR (via ZIP)), documents (HTML, PDF, RTF, DOCX, XSLX, PPTX (via ZIP)), images (GIF, PNG, JPG, JP2 (via MP4), TIFF, Photoshop, DICOM), video (MP4, MKV), audio (MP3 (via ID3v2 header), WAV (via RIFF container)), archives (GZip, TAR).}

\paragraph{Scattered Blind Spots}
Some formats have no clearly abusable range or relocatable information. In this case, the abusable data only comes from some deprecated fields or descriptive text data, such as sections/segment names, title, or used application. Inserting early null characters in fixed fields to abuse the rest of the field is a common approach.

Overall, a vast majority of the common binary file formats can be abused, and room can be made even in arbitrary files for more than 1000 bytes in the first 2048 bytes while retaining the original file functionality.

% \textbf{JavaScript (.js)} files are plain text documents containing JavaScript code. The entire file content is editable, including function definitions, variable declarations, and comments. Care must be taken to maintain syntactic correctness when modifying these files. However, multiline comments can be inserted and as long as no closing comment character is inserted, any arbitrary content can be inserted.

\paragraph{Format Preserving Adversarial Examples}

In general we create a custom process for each different file type that gets an unmodified file as input and return a version of the file that is functionally equivalent and a list of bytes that can be modified. After having a few free bytes in each file format we now measure the success rate of our attack on this setting.
% Note that docx and xlsx are basically the same format: Office Open XML (pptx works too).
We focus on seven commonly used file formats (PDF, ZIP, Docx, Xlsx, ELF, PNG and Javascript) to have coverage over different approaches and file types used to deploy malware. 

Figure~\ref{fig:magika_format} summarizes our attack success rate across each of the seven file types. 
Each of these attacks preserve the original file formats, and their original (malicious) functionality remains valid. 
Specifically, this means we maintain the necessary header for all formats, and modifications are limited to non-critical portions such as code comments in JavaScript and PDF, file names in ZIP files, editing extra jumps in ELF files, Text chunks in PNGs, and comments and names in Office files. 
Office files required the most modifications compared to other formats. However, all file formats demonstrated a high degree of vulnerability and were easily affected. Despite these challenges we achieve a 100\% attack success rate for all file types. If we constrain our edits to just 13 bytes, we can achieve 90\% on all file types.

%\kurt{You'll need to cite why you believe these are the most common malicious sources. Or, explain Google shared these details with you to focus your proof-of-concept attack.}\milad{rewritten to not try to mention google again}

%\ange{PE is by far the most used format for malware deployment, PNG is rarely malicious (see commented out stats)

% Stats on Google Threat Intelligence per file types with 5+ detections:
% search for `type:peexe positives:5+`
% - Documents: PDF:7M, docx:227k, xlsx:132k
% - Executables: PE:157M, android:4M, Mach-o:233k, ELF:2m, jar:135K
% - Images: PNG:4k, Gif:52k, JPEG:4k
%}
%

% \begin{table}[h]
%     \centering
%     \begin{tabular}{llr}
%         \toprule
%         & Task & Cost \\
%         \midrule
%         \multirow{3}{*}{\rotatebox[origin=c]{90}{1-time}} & Recreate training dataset & TODO cost \\
%         & Train transfer models & TODO cost \\
%         & TODO other things & TODO cost \\
%         \midrule
%         \multicolumn{2}{l}{Construct adversarial sample} & TODO cost \\
%         \bottomrule
%     \end{tabular}
%     \caption{There is both an up-front cost to generate adversarial samples and an ongoing cost for each sample.}
%     \label{tab:adversarial_costs}
% \end{table}

\subsection{An end-to-end attack on GMail}

%\magika has been utilized in various Google products, including GMail. 
To demonstrate the potential impact of adversarial attacks against \magika, we conduct an end-to-end experiment showing that it is possible to transmit malware via Gmail under certain conditions.

While our study focuses specifically on Magika’s role in Gmail’s attachment scanning pipeline, our observations suggest that Gmail likely employs a defense-in-depth strategy with additional filtering mechanisms beyond Magika. In particular, we observed that rule-based checks on file extensions influence the routing and filtering of attachments. For example, attempting to upload a file named sample.exe is immediately blocked by the Gmail user interface, preventing it from ever reaching Magika. Similarly, a malicious PDF that retains its .pdf extension appears to be routed to a dedicated PDF malware scanner, bypassing Magika’s classification entirely. To allow our proof-of-concept attack to reach Magika, we removed or altered file extensions, thereby avoiding both hard blocking (as with .exe) and forced routing to specialized scanners (as with .pdf).

For the attack itself, we selected a malicious PDF sample from VirusTotal~\cite{vt} that was not detected by popular antivirus scanners (to avoid triggering signature-based detection). We then applied our adversarial perturbation, additionally modifying the file to evade traditional signature-based file-type detection tools (e.g., \code{file}), and removed the .pdf extension before uploading the attachment to Gmail.

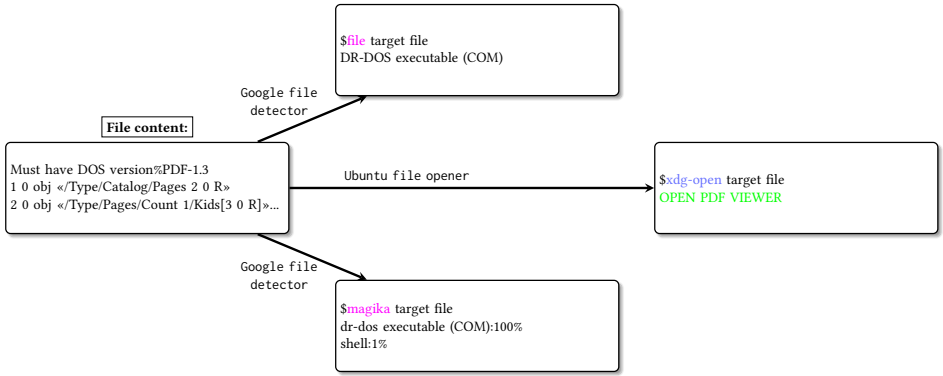
\begin{figure*}
    \centering
    \resizebox{0.7\textwidth}{!}{
    \begin{tikzpicture}[
        node distance = 3cm,
        box/.style = {
            rectangle,
            rounded corners=3pt,
            draw=black,
            thick,
            fill=white!95,
            text width=6cm,
            align=left,
            minimum height=2cm,
            blur shadow={shadow blur steps=5},
        },
        terminal/.style = {
            font=\ttfamily
        },
        arrow/.style = {
            thick,
            ->,
            >=stealth,
            line width=1.5pt
        }
    ]
        % Main PDF text node
        \node[title] (title) {File content:};
        \node[box, below=0.1cm of title] (pdf) {
            \textcolor{black}{Must have DOS version\%PDF-1.3}\\
            \textcolor{black}{1 0 obj <</Type/Catalog/Pages 2 0 R>>}\\
            \textcolor{black}{2 0 obj <</Type/Pages/Count 1/Kids[3 0 R]>>...}
        };
        
        % DR-DOS node with terminal styling
        \node[box, above right=1cm and 1cm of pdf] (drdos) {
            \textcolor{black}{\$}\textcolor{terminalpurple}{file }\textcolor{black}{target file}\\
            \textcolor{black}{ DR-DOS executable (COM)}\\
        };
        
        \node[box, below right=1cm and 1cm of pdf] (magika) {
            \textcolor{black}{\$}\textcolor{terminalpurple}{magika }\textcolor{black}{target file}\\
            \textcolor{black}{ dr-dos executable (COM):100\%}\\
            \textcolor{black}{ shell:1\%}\\
        };
        
        \node[box, right=8cm of pdf] (xdg) {
            \textcolor{black}{\$}\textcolor{terminalblue}{xdg-open }\textcolor{black}{target file}\\
            \textcolor{terminalgreen}{ OPEN PDF VIEWER }\\
            
        };

        % % Magika node with terminal styling
        % \node[box, below right=2cm and -1cm of pdf] (magika) {
        %     \textcolor{terminalgreen}{Magika -r examples/}\\
        %     \textcolor{white}{(detection result)}
        % };
        
        % Draw arrows
        \draw[arrow] (pdf) -- (drdos);
        \draw[arrow] (pdf) -- (magika);
        \draw[arrow] (pdf) -- (xdg);
        
        % Add command labels with terminal styling
        \node[terminal, text width=3cm, align=center]  at ($(pdf)!0.4!(drdos) + (0,0.7)$)  
            {\textcolor{black}{Google file detector}};
        \node[terminal, text width=3cm, align=center] at ($(pdf)!0.4!(magika) + (0,-0.7)$) 
            {\textcolor{black}{Google file detector}};
            
        \node[terminal, text width=3cm, align=center] at ($(pdf)!0.4!(xdg)$) 
            {\vspace{0.5cm}\textcolor{black}{Ubuntu file opener}};
        
    \end{tikzpicture}}
    \caption{A proof of concept for the attack that can be used in GMail to launch an end to end attack. Given the different in how different operating system select what applications to open the files }
    \label{fig:poc}
\end{figure*}

Then, we found bytes of the PDF that could be altered without corrupting the file format, and modified them in such a way that \magika no longer classifies the file as a PDF. 
When we again sent the modified malicious PDF through GMail,
we observed that the message was successfully delivered to a second account under our control (see Figure~\ref{fig:poc}). 
Our results demonstrate that by bypassing \magika and other file-type detection techniques, we are able to successfully route our malicious files to an invalid scanner, defeating protections.

One potential limitation with this proof of concept is that after the victim receives our malicious PDF, the lack of a file extension might impact whether their device will still open the file.
Fortunately for the attacker, depending on a victim's operating system, we can design our PDF such that the process that decides what application to open the unknown file with---such as \code{gnome-open} or \code{xdg-open}---still operates correctly.
This is feasible because the logic used by an operating system differs from the file-type detection used by GMail. Figure~\ref{fig:poc} shows a malicious PDF that can successfully evade GMail's security scanner, but that will correctly open with the PDF viewer on Ubuntu 22.
Alternatively,
for other cases,
the adversary could attempt to social engineer the victim to add the missing extension back to the file
or ask them to execute it directly.

%This attack is in particular easily do able since this tool has been fully open sourced. In the Section~\ref{sec:defense} we will discuss several defense approaches and their effectiveness for our attacks.

\section{Defenses Approaches}\label{sec:defense}

Given the vulnerability of \gmail to white-box adversarial example exploitation methods,
we now take steps to develop defenses that mitigate the efficacy of
this attack.
Specifically, we aim to answer an entirely practical question:
is it possible to increase the difficulty of exploiting the \magika router component of the
defense so that it is no longer the weakest defense component?
Instead of aiming to completely solve the problem of adversarial examples
(something that the research community has been unable to achieve despite
over a decade of research),
our lower bar is both more practical and more realistic.

%\kurt{In general, the point should be can you easily harden a system not originally designed for security; if you can, then your approach can be replicated ideally on all the components, increasing the cost of attacks.}

Towards that end, in this section we design and evaluate defense strategies
that operate under the black-box \emph{transferability} threat model.
In collaboration with \gmail developers,
we evaluate several strategies where the parameters of the
production system are set differently than those of the open-source release.
We investigate 
how various standard defense techniques like randomization would apply 
to this malware classification setting,
but ultimately find none are effective at significantly decreasing the 
attack success rate.
As such, we develop our own defense strategy and
show it significantly reduces the vulnerability of the production
system to transfer attacks.

%
%
%We investigated the vulnerability of the \magika model and demonstrated its susceptibility to attacks. However, our threat model assumed white-box access, which is currently feasible due to Google's public release of the entire model. Given the practical implications of this evaluation, a crucial question arises: how would the situation change if we lacked direct access to the model?
% What are the most effective approaches for releasing open-source models without compromising security? In this section, we will explore basic defense strategies that could limit risks and evaluate their effectiveness. While it has been shown previously that such basic defense approaches cannot prevent adversarial attack, we are facing with much more limited setting of the attack. Moreover, in practical settings we even increasing the cost the attack can in cases be very useful, therefore, we evaluate simpler defenses to get a better understanding of risk of open sourcing different variation of a model. 

\paragraph{Threat Model} 
We adopt a standard \emph{transfer attack} setting from the adversarial ML literature.
The defender deploys a proprietary model $f_{\theta}^\text{target}$ trained on internal data and does not release its parameters or weights.
The attacker does not have direct access to $f_{\theta}^\text{target}$, but can:
(1) access the open-source Magika architecture and training code, 
(2) collect data drawn from a \emph{similar distribution} as the defender’s training data (but not the exact dataset), and 
(3) train one or more \emph{substitute models} $f_{\theta'}$ using this public information and sufficient compute resources.
The attacker crafts adversarial perturbations against $f_{\theta'}$ and attempts to transfer them to $f_{\theta}^\text{target}$.
For each defense we will describe the approach used to train the transfer models. We also assume the adversary has complete knowledge of the algorithm used in the training except the explicit randomness or private secrets of the training.

\subsection{Different Randomization}
\begin{figure}
    \centering
    \includegraphics[scale=0.58]{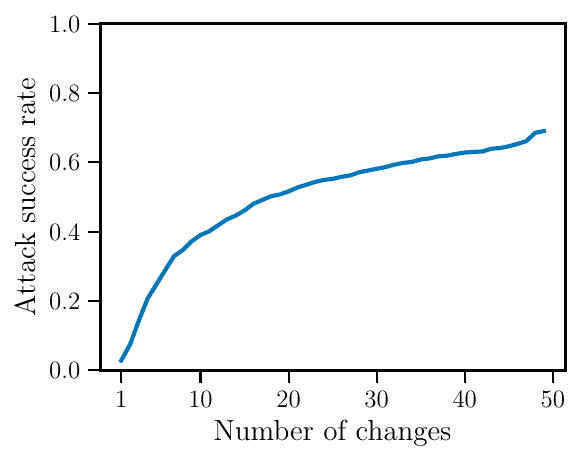}
    \caption{Even if the adversary only has access to a model trained with a different seed, adversarial examples generated on that single model still transfer to the target model. While we observe a decrease in transferability compared to Figure~\ref{fig:direct_magika_results}, the attack remains effective.}
    \label{fig:direct_transfer}
\end{figure}

\begin{figure}
    \centering
    \includegraphics[scale=0.6]{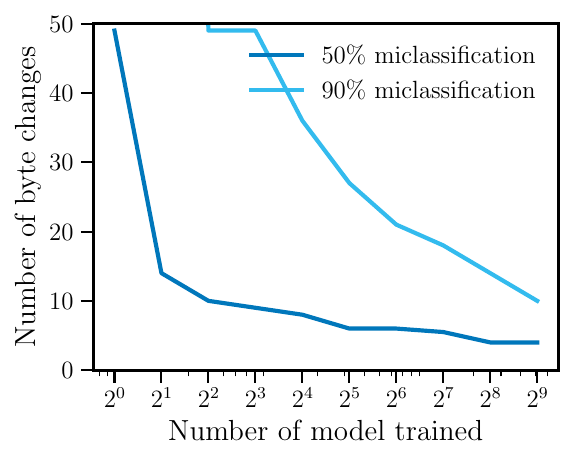}
    \caption{The adversary can improve the transferability of their attack by training additional models. Our results demonstrate that training more models allows the adversary to significantly improve the transfer rate, approaching the effectiveness of an attack directly on the target model.}
    \label{fig:transfer_compute_diff_seed}
\end{figure}

Perhaps the most straightforward defense idea is simply to release a retrained model instead of the original target model.
Because training machine learning models involves a large degree of
randomness, this retrained model will have different learned parameters,
and attacks crafted on the open-source model may not directly transfer
to the closed-source model.

We evaluate this strategy on the dataset described in Section~\ref{sec:exp_setup}, using the same 1,130 samples across all experiments for consistency.
An adversary who constructs evasive malware samples on one such substitute model and transfers them to another succeeds only 35\% of the time with 10 bytes changed---a significant decrease from the 90\% observed in the white-box setting.
Figure~\ref{fig:direct_transfer} shows the complete distortion-versus-success-rate curve.

But an adversary can do better.
Prior work \cite{papernot2016transferability,liu2016delving} has found that one of the simplest and most
effective strategies to produce transferable adversarial examples is
to generate an input that fools multiple white-box models
simultaneously, and then attempts to transfer the attack to the
(unknown) remote model.
Figure~\ref{fig:transfer_compute_diff_seed} illustrates a scenario where the adversary trains multiple models with varying seeds on the data distribution. Adversarial examples are then generated based on a combination of these models, with a final check against the target model.
While a small number of similar models requires substantial changes to evade classification, increasing the number of trained models dramatically simplifies the problem. Interestingly, even a single additional model exhibits significant transferability. Therefore if we use this approach, even adversaries without extensive compute resources can bypass detection by either making larger file modifications or attempting multiple smaller modifications.

\subsection{Different Model Architecture}

\begin{figure}
    \centering
    \includegraphics[scale=0.6]{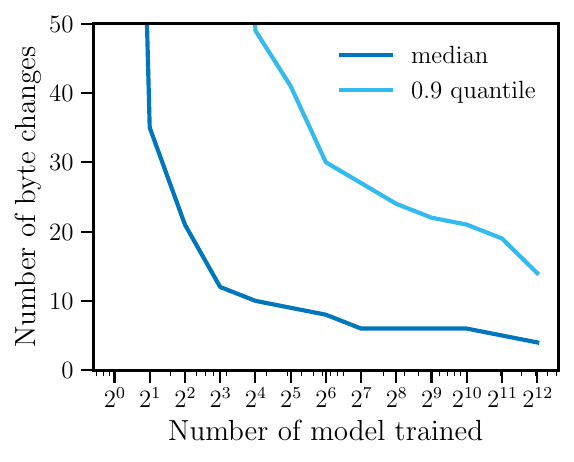}
    \caption{By training multiple models, an adversary can reduce the size of the necessary perturbation. Shown are the number of bytes an adversary must change to maintain a 50\% and 90\% attack success rate.}
    \label{fig:trasnfer_compute_diff_archs}
\end{figure}
\begin{figure}[t]
    \centering
    \includegraphics[scale=0.6]{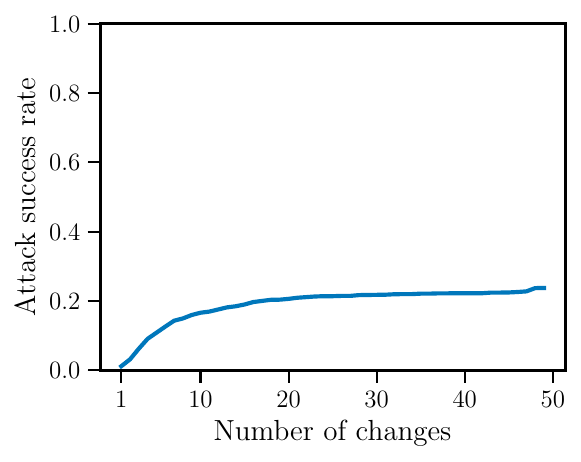}
    \caption{When the model trainer employs different model architectures for the released model compared to the target model, an adversary can still discover examples that transfer between the two. However, the success rate of such transfers is notably reduced in contrast to situations where the model architectures are the same.}
    \label{fig:direct_transfer_diff_arch}
\end{figure}

Although adversarial examples can be transferred between models, adversaries often have limited computational resources.
We now evaluate the effectiveness of transfer attacks under a constrained-compute setting.
Figure~\ref{fig:direct_transfer_diff_arch} shows that if the adversary has access only to a single substitute model released by the model trainer, more modifications to the file are necessary to evade the target classifier.

To strengthen the attack under the same compute budget, the adversary can train an \emph{ensemble of substitute models} with \emph{diverse architectures} and search for perturbations that fool all of them simultaneously.
In this setting, architectural diversity is introduced by varying the convolutional depth, kernel sizes, and layer widths of the Magika architecture (details in Appendix~\ref{app:architectures}).
We train $N=K$ such models and generate adversarial examples that optimize the combined loss across the ensemble before testing transferability to the target model.
Figure~\ref{fig:trasnfer_compute_diff_archs} summarizes the results, showing that increasing the number and diversity of substitute models improves transfer success compared to the single-model baseline (Figure~\ref{fig:transfer_compute_diff_seed}), though even this strategy cannot guarantee evasion against a defender with substantial computational resources.

\subsection{Different Data Pre-processing}

Deep learning models are known for their ability to learn directly from raw data, however, in many cases, such as the \magika system, model trainers make specific choices about how the data is represented or preprocessed. In \magika, for example, only certain subsets of bytes from the original file are used as input to the model. Therefore we can consider a setting where the released model and the target model use different data processing approaches. We aim to assess how effective this approach is in protecting against attacks. 
\begin{figure}[t]
    \centering
    \includegraphics[scale=0.6]{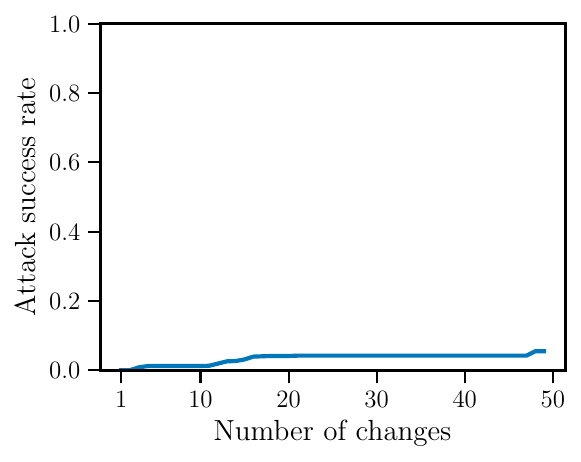}
    \caption{If the model trainer uses different data processing approaches, it significantly increases the difficulty for an adversary to directly transfer attacks between the released model and the target model.}
    \label{fig:direct_transfer_diffoffset}
\end{figure}

We consider a scenario where the defender deploys a target model that follows the Magika architecture~\cite{fratantonio2024magika} but uses modified preprocessing offsets when selecting the input bytes. Specifically, the model extracts the first $b_1$ bytes starting at offset $o_1$, a middle segment of $b_2$ bytes starting at $o_2$, and the last $b_3$ bytes starting at $o_3$, such that $b_1 + b_2 + b_3 = 1536$, the total number of bytes used by the released model. These offsets $(o_1,o_2,o_3)$ are unknown to the attacker.

\begin{figure}
    \centering
    \includegraphics[scale=0.58]{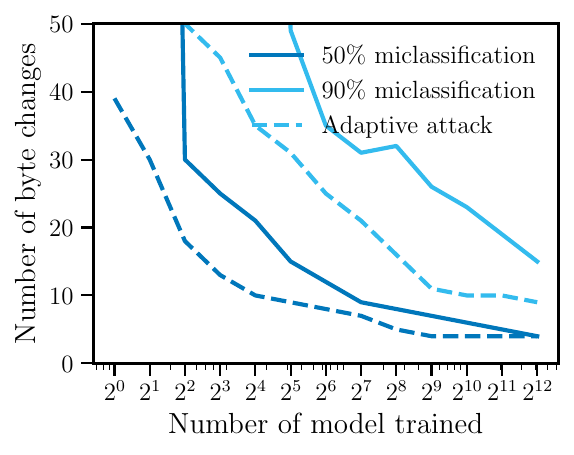}
    \caption{By training multiple transfer models, adversaries can increase the attack success rate to above 90\%. %While employing different data processing approaches can significantly reduce the transferability of attacks, adversaries can still train multiple models using varied data processing approaches to enhance their attacks. As shown in this figure, this requires a greater number of trained models compared to previous approaches, however it still can be an effective method to create transferable adversarial examples.
    In the baseline attack, the adversary trains using random hyperparmaeters; in the adaptive case, the adversary first identifies the best hyper-parameters used in data processing of the target model and then trains additional models with these hyperparameters.}
    \label{fig:trasnfer_compute_diff_offset}
\end{figure}

To study transferability under this defense, we trained substitute models where offsets and byte counts were \emph{randomly sampled} from a search space that shifted each segment by up to a few hundred bytes in either direction while preserving the 1,536-byte total. Figure~\ref{fig:direct_transfer_diffoffset} shows that this mismatch between preprocessing pipelines significantly lowers transfer success compared to the baseline setting, when only one single model used for the transfer attacks. Then, we trained ensembles of up to $2^12$ such variants and optimized adversarial examples against their combined loss before evaluating on the defender's model with its hidden offsets. As we can see in Figure~\ref{fig:trasnfer_compute_diff_offset} as we increase the number of the models in the ensemble, the attack success rate also increases. Therefore, a well resourced adversary can defeat this defense.

We further consider an \emph{adaptive attack} in which the adversary knows the general slicing algorithm but not the exact offsets. The attacker sends probe files with varied synthetic byte shifts and observes the target's classification outcomes, allowing them to narrow down likely offsets. New substitute models are then trained on these candidate offsets, and adversarial perturbations are recomputed across this refined ensemble. As shown by the dashed lines in Figure~\ref{fig:trasnfer_compute_diff_offset}, adaptively searching offsets reduces the compute required to find transferable examples but still incurs a higher cost than the standard transfer scenario.

A key advantage of basic defense approaches is that they generally maintain the model's performance on clean input. This is crucial in security-sensitive applications where high false positive rates can significantly impact usability for normal users. In the next section, we will explore additional defense strategies that involve modifying the training process or model architecture, which may affect the overall accuracy.

% \subsection{Different training data}

% \section{Advance Defenses}\label{sec:defense_adv}

\subsection{Adversarial Training}

\begin{figure}[t]
    \centering
    \includegraphics[scale=0.6]{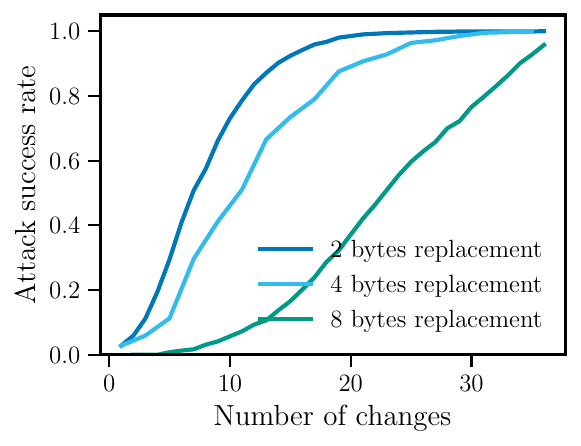}
    \caption{As we increase the perturbation bound of the attack used during adversarial training, the robustness of the resulting model increases. However, in all cases, once the attacker applies a perturbation greater than what was used during training, the attacks begin to succeed with near-100\% success rate.}
    % \vspace*{-0.7cm}
    \label{fig:trasnfer_adv_training}
\end{figure}

\begin{figure}
    \centering
    \includegraphics[scale=0.6]{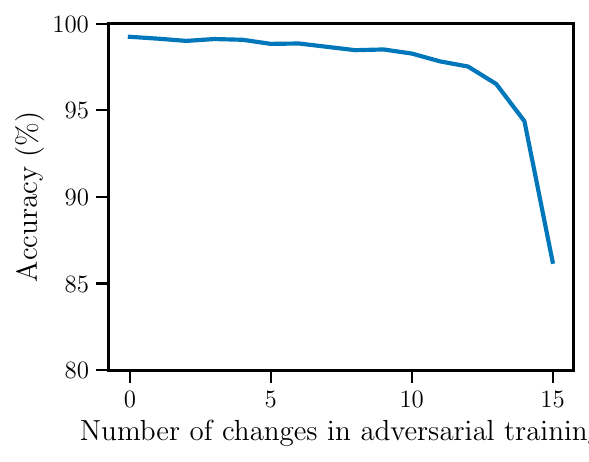}
    \caption{ Increasing perturbation bound of the attack used during adversarial training reduces the accuracy of the classifier on benign data. At a perturbation bound of 0 (i.e., no attack) the accuracy is 99\%; reducing this to an attack that perturbs 10 bytes marginally decreases the benign accuracy to 97\%, while changing 15 bytes significantly degrades the classifier's utility to below 90\% accuracy.}
    \label{fig:trasnfer_adv_training_utility}
\end{figure}

The most successful adversarial example defense is adversarial training~\cite{madry2017towards}, which assumes the defender knows the family of attacks that the adversary will employ and can train a model robust to these specific perturbations.
For our evaluation, we reproduce this defense in the Magika setting by training a second "private" model using the same architecture as the public model, augmented with adversarially perturbed examples generated during training.

We generate adversarial examples on-the-fly in each mini-batch using our modified GCG attack (Algorithm~\ref{alg:gcg_blindspot}) applied to the public released model.
At each training step, for a clean sample $x$, we compute the gradient of the loss with respect to the input bytes and greedily select coordinates to modify, one at a time, up to a perturbation budget of $k \in \{5,10,15\}$ bytes.
For each coordinate, we evaluate candidate byte substitutions and choose the one that maximally increases the cross-entropy loss on the public model.
The resulting adversarial example $x'$ is paired with its ground-truth label and added to the batch alongside clean samples.
The private model is then trained to correctly classify both $x$ and $x'$ by minimizing the standard cross-entropy loss on this mixed batch. We use up to 100 epoch of training and select the checkpoint with the highest average accuracy on the test set.

Figure~\ref{fig:trasnfer_adv_training} shows that this approach improves robustness for low-byte attacks (e.g., $k=5$), but as the allowed number of modified bytes increases, adversarial examples still transfer successfully.
The clean accuracy of the private model drops from $99\%$ without adversarial examples ($k=0$) to $97\%$ for $k=10$, and below $90\%$ for $k=15$ (Figure~\ref{fig:trasnfer_adv_training_utility}).
This degradation highlights a fundamental trade-off: stronger adversarial training reduces utility in large-scale classification settings where even a small drop in accuracy impacts millions of users.
Moreover, a determined adversary can replicate the same adversarial training strategy on their own substitutes, reducing the defense’s long-term effectiveness.
We therefore do not further pursue this defense beyond this baseline study.

\subsection{Similarity Unpairing} % Adjust \subsection if needed

While adversarial training can offer robustness against specific known attacks, it often impacts benign accuracy and may not generalize well against unforeseen attack variations. Another approach for defense is  aimed to reduce transfer attacks by introducing differences (randomization, architecture, preprocessing) between a potentially public model and the target deployment model as explored before. However, as shown, determined adversaries can often overcome these differences by training multiple substitute models.

Hong et al.,~\cite{hong2023publishing} argue that the main vulnerability enabling transfer attacks is the potential \textit{similarity} in the input-gradient landscape between the substitute model(s) available to the attacker (analogous to an open-source release or $f_{\theta_s}$ in a general scenario) and the actual deployed target model ($f_{\theta_o}$). If gradients point in similar directions for similar inputs, perturbations crafted on one model are more likely to affect the other.

We explore \textbf{Similarity Unpairing}~\cite{hong2023publishing} as defense mechanism for the file classification system. The intuition is to explicitly train the public model ($f_{\theta'}$) such that while maintaining high accuracy similar to a private model ($f_{\theta}$), its input-gradient landscape is intentionally made dissimilar to $f_{\theta}$. This dissimilarity aims to directly disrupt the effectiveness of gradient-based transfer attacks.

Formally, starting from a trained model $f_{\theta}$, we fine-tune it to obtain the public model $f_{\theta'}$ using a modified objective function:
\begin{equation}
    \mathcal{L} = \mathcal{L}_{xe}(f_{\theta'}(x), y) + \lambda \cdot \mathcal{C}_s(x, y, f_{\theta}, f_{\theta'})
\end{equation}
where, $\mathcal{L}_{xe}$ represents the cross-entropy loss ensuring the model remains accurate for the primary classification task. $\mathcal{C}_s$ is a similarity function designed to measure the input-gradient similarity between the original model $f_{\theta}$ and the fine-tuned model $f_{\theta'}$. The hyperparameter $\lambda$ balances the standard accuracy objective with the goal of minimizing gradient similarity. By minimizing $\mathcal{C}_s$ during fine-tuning, the approach "unpairs" the gradient landscapes of the two models.

For our settings adapting this approach, we would fine-tune the public model for a small number of epochs (10 epochs) using this combined loss. The specific definition of the similarity function $\mathcal{C}_s$ can vary; we use cosine similarity between the gradients of the two models with respect to the input $x$. 
 
\begin{figure}[t]
    \centering
    \includegraphics[scale=0.6]{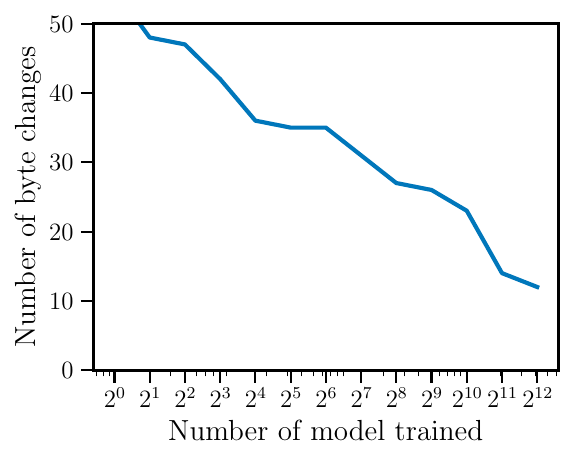}
    \caption{Increasing the number of additional model trained by unpairing approach can increase the transferability of adversarial examples.}
    % \vspace*{-0.3cm}
    \label{fig:trasnfer_unpairing_training}
\end{figure}

In the setting we consider, the adversary is aware that the defender is using the similarity unpairing approach. As an attack, the adversary trains substitute models using the same unpairing loss function to create different fine-tuned versions of the released model. First, we observed that this defense approach does not significantly affect the accuracy of the released model. Figure~\ref{fig:trasnfer_unpairing_training} summarizes the results for this defense against transfer attacks. As can be seen, without any additional substitute models trained by the adversary, the defense reduces the effectiveness of the transfer attack. However, as the adversary increases the number of additional models trained with unpairing, the attack effectiveness increases, similar to the previous settings; we observe a large drop in the number of bytes required to achieve a misclassification with a $90\%$ success rate. Therefore, while this defense provides some benefit in limited settings, a resourceful adversary can likely bypass it by training additional models.

\subsection{Our Proposed Defense}
One challenge in defending against adversarial examples is that attackers can often estimate the preprocessing steps used and craft attacks that transfer effectively across different models.  While using diverse preprocessing pipelines can help, the limited space of common preprocessing choices makes it feasible for attackers to approximate these transformations. Moreover, we saw that in the previous defense when we try to unpair the similarity of the input spaces of the model, the transfer rate is reduced.  We combine the ideas from both defenses and we propose a novel defense inspired by cryptographic techniques, specifically the Advanced Encryption Standard (AES)~\cite{dworkin2001advanced}.

AES is a symmetric block cipher that encrypts data using a series of substitution, permutation, and mixing operations~\cite{dworkin2001advanced}.  These operations are organized into rounds, and the number of rounds depends on the desired security level.  A key feature of AES is its strong diffusion and confusion properties. Diffusion ensures that even small changes in the input propagate throughout the encryption process, resulting in significant changes in the output. Confusion obscures the relationship between the input and output, making it difficult for attackers to analyze the cipher. In our defense, we leverage these properties by incorporating a modified AES transformation into the preprocessing pipeline. However, using the full AES encryption process would make the model overly sensitive to even single-bit perturbations in the input. Therefore, we apply only a limited number of AES rounds. Our experiments demonstrate that even a single round of AES provides significant protection against transfer attacks, effectively disrupting the attacker's ability to estimate the preprocessing pipeline and generate transferable adversarial examples. This approach introduces a dynamic and non-linear transformation that significantly expands the space of possible preprocessing functions, making it much harder for adversaries to approximate.

\begin{algorithm}[t]
\caption{AES-Based Preprocessing Algorithm (ECB)}\label{alg:aes}
% \footnotesize
\begin{algorithmic}[1]
\Require Input data bytes $D$, AES encryption key $K$, Number of AES rounds $R$
\Statex \textbf{Splitting:}
\State $n \gets \text{length of } D \text{ in bytes}$
\State $p \gets n \mod 16$  \Comment{Calculate padding needed}
\If{$p \neq 0$}
    \State $D \gets D \mathbin\Vert 0^{16-p}$  \Comment{Pad $D$ with zeros to a multiple of 16 bytes}
\EndIf
\State $N \gets n \div 16$  \Comment{Number of 16-byte blocks}

\Statex \textbf{AES Encryption (ECB Mode):}
\For{$i = 1 \text{ to } N$} 
    \State $B_i \gets D_{(i-1) \times 16 + 1 \text{ to } i \times 16}$  \Comment{Extract the $i$-th 16-byte block}
    \State $B_i \gets \text{AES-Encrypt}(B_i, K, R)$  \Comment{Encrypt $B_i$ using key $K$} 
    \State $D'_{(i-1) \times 16 + 1 \text{ to } i \times 16} \gets B_i$  \Comment{Store the encrypted block back into $D'$}
\EndFor

\Return $D'$
\end{algorithmic}
\end{algorithm}

Algorithm~\ref{alg:aes} outlines our input preprocessing approach. Because the AES block size is 16 bytes and our model's input size exceeds this, we split the input into multiple 16-byte blocks.  AES has several modes of operation, including ECB (Electronic Codebook) and CBC (Cipher Block Chaining)~\cite{dworkin2001advanced}. We report only the results for ECB mode in the paper; we observe that CBC model has high effect on the utility of the model.

\begin{figure}[t]
    \centering
    \includegraphics[scale=0.55]{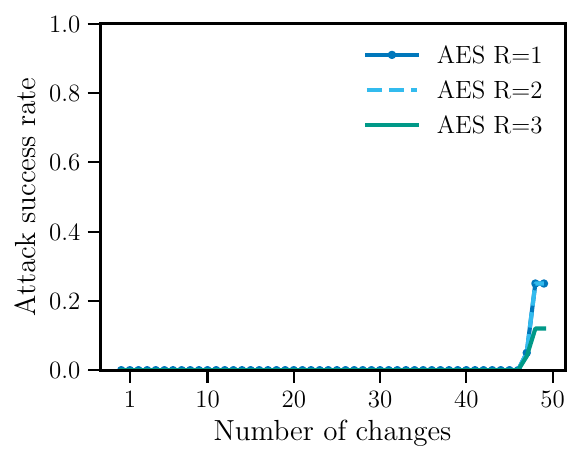}
    \caption{Our proposed defense is highly effective in reducing the transfer rate between two models trained with different keys.}
    \label{fig:aes}
    \vspace*{-0.1cm}
\end{figure}

\begin{figure}[t]
    \centering
    \includegraphics[scale=0.55]{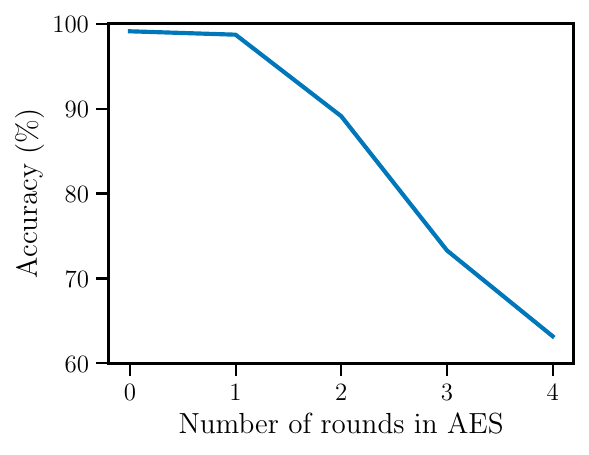}
    \caption{Increasing the rounds AES significantly reduces the utility of the final model.}
    \label{fig:aes_acc}
    \vspace*{-0.7cm}
\end{figure}

Our results demonstrate the effectiveness of incorporating AES encryption into the preprocessing pipeline as a defense against transfer attacks. First we show our approach does not heavily affect the utility of the model: as we can see in Figure~\ref{fig:aes_acc}, when we increase the number of round of AES, we see a significant drop in the accuracy. Nevertheless,  we can get effective results even with one round of AES. 

Figure \ref{fig:aes} shows that our proposed defense significantly reduces the transferability of adversarial examples. In this experiment, the defender uses only one round and the adversary trains 512 models with different keys and seeds for training on the similar underlying dataset (different sampled from the large training data) and generate adversarial examples that fool all of them at the same time and then transfer them to the larger models. When we have a large number of modifications we can still see some misclassifications, however, this might be due to classification error and not transferability of adversarial examples . In Appendix~\ref{app:add_res} we discuss the adversary who can directly attack the target model and not leverage transfer attack and achieve a higher success rate than using transfer attack.

It is important to emphasize that we do not claim to defend against all transferable adversarial examples. Attacks generated from entirely unrelated models might still transfer to our target model. Our focus here is specifically on how releasing an open-source version of a model can increase the attack surface. Our findings demonstrate that if the private version of a model employs a defense similar to ours, an adversary with access to the open-source version gains no advantage in crafting successful attacks.

%We shared the results of this defense analysis with the \gmail
%developers who indicated their intent to apply the defense we suggest to the production %classifier.
We finally implement this defense and, in the collaboration with Google engineers, we deploy it within the production \gmail classifier
in order to improve user safety.
As such, the attacks introduced in the prior section are unlikely to remain effective
after the release of this paper.

\paragraph{Latency trade-off.}
In addition to classification utility, we evaluated the computational overhead introduced by the AES-based preprocessing. 
Our implementation processes the 1,536-byte input in 16-byte blocks using a vectorized AES routine optimized for CPU inference. 
With a single AES round (the setting we use for deployment), the additional latency per file is below a microsecond on standard production hardware, which is negligible compared to the overall inference time of the Magika pipeline. 
Even when increasing the number of rounds to three, the total added latency remains well under one millisecond per file, making the defense practical for large-scale, high-throughput malware scanning systems.

\begin{figure}
    \centering
    \includegraphics[scale=0.55]{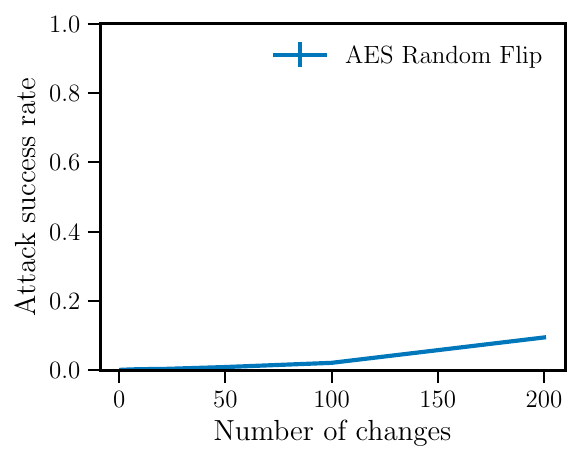}
    \caption{Our proposed approach can still fail when the adversary can change large number of bytes even at random! (averaged between 5  runs) }
    \label{fig:aes_random_flip}
\end{figure}

\section{Limitations}
Our evaluation focuses on transfer attacks, where the adversary does not have access to the parameters or secret key of the deployed model. In this setting, the proposed AES-based preprocessing significantly reduces transfer attacks. However, a white-box adversary with full system access could still succeed. In the Gmail proof-of-concept, a successful attack required both altering or removing the file extension and bypassing Magika as well as other rule-based classifiers. This step is specific to Gmail’s defense-in-depth pipeline and may reduce effectiveness in practice, since file handling behavior varies across operating systems and some users may be less likely to open files without familiar extensions (see Figure~\ref{fig:aes_random_flip}).

We again reiterate an important limitation of our work: we report attacker cost in terms of the number of bytes modified, but acknowledge this does not fully capture real-world effort—motivated adversaries could modify more bytes if needed. Also, malicious functionality was verified for the subset of files used in our proof-of-concept in isolated environments, but not exhaustively across all formats. Finally, while the defense is designed to reduce transferability from public to private models, it can still be bypassed by large-budget black-box search or direct white-box attacks.

%\kurt{One thing worth summarizing at the end of all of these experiments: are there meaningful changes that actually make this robust, or is it all-together trivial for an attacker to evade, even after hardening techniques? A 50-byte change with the AES scheme still feels simple for an attacker to mount.}
%\luca{A potential metric to evaluate kurt's point is to measure how many tries the attacker has to make to get a malware through Gmail's defenses. If this number is >> the 99\% percentile of uploads per day for normal users, or if enough evidence is collected to detect the attack with a specialized model, then we have a useful defense. Before the defense, it sounds like a single upload was all that was needed. }

\section{Future work}\vspace{-0.1cm}
This work opens up several promising directions for future research. 
We believe it would be interesting to refine and simplify our defense mechanisms for our task by leveraging domain specific knowledge about different file types. For example, this could involve developing specialized adversarial training approaches that exploit the unique vulnerabilities and blind spots of each file type. Alternatively, we could explore how to design  adversarial training pipelines that exploit the specific characteristics of PDF files or images, and then train models to be robust against these attacks.

We also believe it would be interesting to investigate the effectiveness of our defense in other domains. Our current hypothesis is that the success of our approach in this context stems from the significant differences in patterns across various file types. By introducing diffusion and confusion through cryptographic techniques, we disrupt the attacker's ability to transfer adversarial examples without significantly impacting the utility of the model. However, it is unclear whether this approach will generalize to harder tasks where the underlying data patterns may be more complex and subtle. Applying similar defenses in such domains could potentially lead to a significant decrease in model utility. Therefore, future work should systematically evaluate the effectiveness and trade-offs of our defense across a range of tasks and datasets.

\section{Discussion and Conclusion}

This paper focuses on a specific technical vulnerability – causing a file-type classifier to misclassify – but its significance lies in the system-level impact: how such a vulnerability in one component can compromise a larger security pipeline.
And we show that, by leveraging recent advances in adversarial machine
learning, it is possible to do this by changing just a few bytes
in any given file.
Fortunately, as we show,
by applying relatively straightforward defense ideas, it becomes
possible to increase the robustness of this classifier to attack
by $75\%$ even in scenarios where the adversary uses large amounts of compute to train additional models using the exact same pipeline.

Beyond this specific technical problem, our broader goal is to highlight the need for practical security analysis of real-world systems incorporating ML components. We aim to encourage research that moves beyond attacking isolated models to understanding how attacks manifest and can be mitigated within complex, deployed environments.
This is because machine learning models are no longer isolated to single-purpose
products that only classify images, or transcribe the speech in an audio file.
Today's machine learning models are sufficiently capable that they can be
embedded into tools that from the outside would not give the impression
there is a machine learning component in the overall system.

And because of this, for any given system,
it becomes increasingly likely that there will exist some machine learning model
as a component of this system.
This means that
instead of imagining possible applications of machine learning models and then 
attacking those hypothetical systems,
we argue that future adversarial
machine learning works should also attempt to look for real systems and then develop
attacks that actually cause some specific harm.
In the case of this paper, we have found that the security of the malware classifier
in \gmail depends, in part, on a single neural network
that is vulnerable to transferable adversarial examples.
But we do not believe this will be the only system
that has this property;
in this way, our analysis can serve as a case study for how to
evaluate the robustness of a security system that happens to contain
a machine learning model and how someone might design and evaluate defenses for such systems.

Overall, we hope to encourage future work to take this same approach, and analyze
end-to-end systems that contain machine learning models as subsystems which
(either implicitly or explicitly) are part of the trusted computing base for
the system. Furthermore, we advocate for the development and evaluation of defenses based on their effectiveness in raising the bar for attackers in practice, contributing to incrementally more secure real-world deployments.

% \newpage

\section*{Ethics Considerations}
We strictly followed common responsible disclosure protocols to ensure that no harm was caused in the development of our attack, and no real users were ever targeted for this work.
Prior to validating our attack was effective over the production Gmail classifier we received advance permission from the Gmail team at Google,
and after we completed our defense analysis we worked directly with the affected product team to develop a solution.  
As a result of our disclosing this attack, none of the attacks described in this work remain effective against the deployed system. 
We hope this research encourages more open sourcing of security research and solutions, demonstrating the positive outcomes of collaborative vulnerability discovery and mitigation. 
We received full approval from the affected team to publish these results.

\section*{Open Artifacts}
We release the code for our defense approach alongside the main \magika GitHub repository at \url{https://github.com/google/magika}. However, due to the proprietary nature of the training data used in this work, we cannot share it directly. Instead, we will provide resources in collecting similar training data. All attacks used in this work are based on existing adversarial attack approaches, which are publicly available.

For ethical reasons, we opt not to publicly release our attack source code to avoid attackers leveraging it as part of real malware campaigns. Universities may reach out for access.

\bibliography{refs}{}
\bibliographystyle{plain}

\appendix

\section{Additional results}
\label{app:add_res}

\begin{figure}[t]
    \centering
    \includegraphics[scale=0.55]{{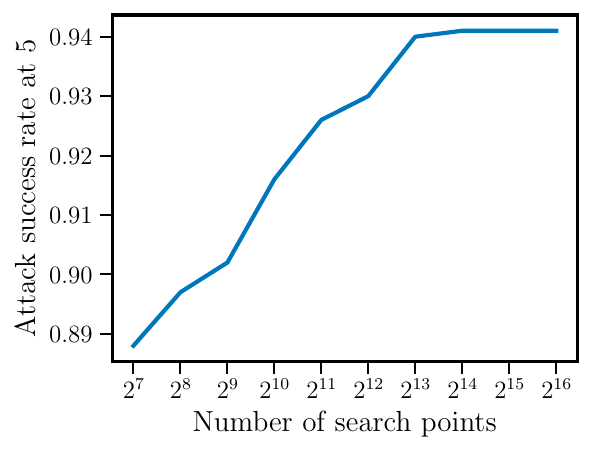}}
    \caption{By increasing the number of neighbors searched in each step of the GCG optimization algorithm,
    we can improve the attack efficacy until we see diminishing returns at $2^{13}$ neighbors searched.}
    \label{fig:top_x}
\end{figure}

\paragraph{Compute vs Attack Success Rate}
The GCG attack has two primary steps: 
first, it identifies the positions and corresponding values that change and then explore various points to find the one that maximizes the adversary objective. The number of points evaluated significantly impacts the attack's effectiveness by increasing computational overhead. Figure~\ref{fig:top_x} illustrates how varying the number of trials per step can influence attack performance. While expanding the search space can enhance attack effectiveness, this improvement is generally limited. The gradient effectively highlights crucial points, and increasing the search size beyond a certain threshold yields diminishing returns.

\paragraph{Real attackers do not compute gradients:} Some researchers argue that in practice, adversaries do not use gradient-based approaches to circumvent machine learning defenses~\cite{apruzzese2023real}. This could also be applicable here, where an adversary might employ fuzzing techniques to measure model accuracy and generate adversarial examples without relying on gradients. We assume such an adversary uses a similar attack strategy as GCG and the main difference is that it randomly selecting neighbors and choosing the best candidate for updates instead of computing gradients. 

% Figure~\ref{fig:random_nolimit} and Figure~\ref{fig:random_flip} demonstrate the effectiveness of this approach, showing that this approach requires many more changes compared to the base. However, there's no compelling reason why an adversary wouldn't utilize gradient-based approaches, especially if they have access to the model. %\kurt{To your earlier point, the number of bytes changed dont really matter. What's interesting is whether your defenses substantially increase the effort to fuzz a sample, versus using gradients.}

\section{Additional experimental details}

\label{app:architectures}

To evaluate ensemble-based transfer (Figures~\ref{fig:direct_transfer_diff_arch}--\ref{fig:trasnfer_compute_diff_archs}), we trained multiple substitute models that preserve Magika's overall design but vary in architectural details. This simulates a realistic attacker with knowledge of the general architecture but not the exact model parameters. All substitutes were trained using Magika's preprocessing, loss function, and data methodology (Section~\ref{sec:exp_setup}).

We achieved architectural diversity by \emph{randomly sampling} variations in convolutional depth (1-4), kernel sizes (8-16), and the number of fully connected layers (1-4). Each substitute architecture was required to achieve at least $95\%$ average test accuracy.

For the attacks, we optimized adversarial examples against the combined loss of an ensemble of these substitute models, simulating a stronger attacker with more compute. We then measured the transfer success rate of these examples on the target model. The number of substitutes and compute budgets are detailed in Section~\ref{sec:exp_setup}.

% \begin{figure}[t]
%     \centering
%     \includegraphics[scale=0.5]{figures/number_changes_blackbox.pdf}
%     \caption{Even without any gradient information an adversary can use black-box optimization technique to evade the classifier.
%     (But using gradient information reduces the number of bytes modified by $4\times$ on average.)}
%     \label{fig:random_nolimit}
% \end{figure}

% \begin{figure}[t]
%     \centering
%     \includegraphics[scale=0.5]{figures/number_changes_topx_blackbox.pdf}
%     \caption{In a black-box setting, where we lack access to gradients to guide optimization, the number of search points becomes crucial. In contrast to Figure~\ref{fig:top_x}, where using gradients eventually leads to diminishing returns from increasing search size, we observe that in the black-box scenario, expanding the search space continues to improve the attack's effectiveness.}
%     \label{fig:random_flip}
% \end{figure}

% \vspace*{-1cm}

\end{document}